\documentclass[fleqn,usenatbib]{mnras}

\usepackage{newtxtext,newtxmath}
\usepackage[dvipsnames]{xcolor}

\usepackage[T1]{fontenc}

\DeclareRobustCommand{\VAN}[3]{#2}
\let\VANthebibliography\thebibliography
\def\thebibliography{\DeclareRobustCommand{\VAN}[3]{##3}\VANthebibliography}

\usepackage{graphicx}	
\usepackage{amsmath}	
\usepackage{subcaption}
\usepackage{pifont}
\usepackage{arydshln}
\usepackage{xcolor}
\usepackage{caption}
\usepackage{dashrule}
\usepackage{fix-cm}

\title[Environmental dependence of galaxy properties]{Insights into the dependence of galaxy properties on the environment with explainable machine learning models}
\author[S. S. Uchida et al.]{
Shun-ya S.\ Uchida$^{1}$ \thanks{E-mail: uchida\_shunya@nagoya-u.jp}
, Suchetha Cooray$^{2,3}$ \thanks{E-mail: cooray@stanford.edu}
, Atsushi J.\ Nishizawa$^{4,5,6}$
, Tsutomu T.\ Takeuchi$^{1,7}$, 
\newauthor  and Peter Behroozi$^{8}$
\\
$^{1}$Graduate School of Science, Nagoya University, Furocho, Chikusa Nagoya 464-8602, Aichi, Japan \\
$^{2}$Kavli Institute for Particle Astrophysics and Cosmology, Stanford University, Stanford, CA 94305, USA \\
$^{3}$SLAC National Accelerator Laboratory, 2575 Sand Hill Road, Menlo Park, CA 94025, USA\\
$^{4}$Digital Transformation (DX) Center, Gifu Shotoku Gakuen University, 1-1 Takakuwanishi, Yanaizu-cho Gifu 501-6194, Gifu, Japan\\
$^{5}$Institute for Advanced Research, Nagoya University, Furocho, Chikusa Nagoya 464-8602, Aichi, Japan\\
$^{6}$Kobayashi-Maskawa Institute for the Origin of Particles and the Universe (KMI), Nagoya University, Furocho, Chikusa Nagoya 464-8602, Aichi, Japan \\
$^{7}$The Research Center for Statistical Machine Learning, The Institute of Statistical Mathematics, 10-3 Midori-cho, Tachikawa 190–8562, Tokyo, Japan \\
$^{8}$Department of Astronomy and Steward Observatory, University of Arizona, Tucson, AZ 85721, USA
}

\date{Accepted XXX. Received YYY; in original form ZZZ}

\pubyear{\the\year{}}

\begin{document}
\label{firstpage}
\pagerange{\pageref{firstpage}--\pageref{lastpage}}
\maketitle

\begin{abstract}
Galaxies reside within dark matter halos, but their properties are influenced not only by their halo properties but also by the surrounding environment.
We construct an interpretable neural network framework to characterize the surrounding environment of galaxies and investigate the extent to which their properties are affected by neighboring galaxies in IllustrisTNG300 data ($z=0$).
Our models predict galaxy properties (stellar mass and star formation rate) given dark matter subhalo properties of both host subhalo and of surrounding galaxies, which serve as an explainable, flexible galaxy-halo connection model.
We find that prediction accuracy peaks when incorporating only the nearest neighboring galaxy for stellar mass prediction, while star formation rate prediction benefits from information from up to the third-nearest neighbor. 
We determine that environmental influence follows a clear hierarchical pattern, with the nearest neighbor providing the dominant contribution that diminishes rapidly with additional neighbors. We confirm that central and satellite galaxies, as well as different galaxy categories based on mass and star-forming activity, exhibit distinct environmental dependencies. Environmental dependence for low-mass galaxies ($\log(M_*/M_\odot) < 10$) shows 35-50\% environmental contribution compared to just 8-30\% for massive centrals, while satellite galaxies experience consistently stronger environmental effects than centrals across all populations. 
Furthermore, we find that the most significant attribute from neighboring subhalos for predicting target galaxy properties is its distance to the nearest neighboring galaxy. 
These quantitative results offer guidance for constructing more sophisticated empirical and semi-analytic models of galaxy formation that explicitly include environmental dependence as a function of galaxy type and mass.
\end{abstract}




\section{Introduction} \label{sec:intro}
The formation and evolution of galaxies are understood within the framework of the $\Lambda$CDM cosmology, hierarchical structure formation, dark matter (DM), and halos that grow together with the galaxies in them \citep{1978MNRAS.183..341W}. Therefore, the properties of galaxies are intimately linked to the characteristics and distribution of their host halos (and subhalos). Understanding this galaxy-halo connection is fundamental to galaxy formation theory (see \citealp{annurev:galaxy-halo_connction} for a review). A robust statistical relationship exists between stellar mass ($M_*$) and halo mass ($M_h$), known as the stellar-to-halo mass relation \citep[SHMR: e.g.,][]{Moster_2010, Behroozi_2010, 2013ApJ...770...57B, 2017MNRAS.470..651R, SHMR}. However, galaxy properties are not determined solely by $M_h$.

Beyond the primary $M_h$ dependence, galaxy-halo connections exhibit a secondary bias on galaxy properties and their spatial distribution, known as assembly bias. Assembly bias suggests that at a given $M_h$, galaxies cluster more strongly for earlier forming halos that also often have higher halo concentrations \citep[e.g.,][]{Wechsler_2002,Sheth_2004,Gao2005,Wang_2007,Dalal_2008,Hahn_2009,Dalal_2008,Croton_2007,Lim2016,Montero2017,Mao_2018,Zehavi_2018,camargo2025}. 

The secondary dependence on the galaxy-halo connection can also be examined through the environment. 
In the local environment scales, galaxies undergo processes that alter their gas condition: ram-pressure stripping \citep[e.g.,][]{ram_1, ram_2}, galaxy harassment \citep[e.g.,][]{mooreGalaxyHarassmentEvolution1996, Lin_2010}, mergers \citep[e.g.,][]{hopkinsUnifiedMergerdrivenModel2006, Lin_2010}, and strangulation \citep[e.g.,][]{strangulation1, pengStrangulationPrimaryMechanism2015}. 
These changes in gas conditions can influence the galaxy's physical properties, such as color and star formation rate (SFR), within a central galaxy's virial radius.
The properties of satellite galaxies are strongly correlated with those of their central galaxy within their halo, a phenomenon known as "one-halo" galactic conformity \citep[e.g.,][]{weinmannPropertiesGalaxyGroups2006,1halo_sim, 1halo_theori}. 
There is also evidence that the environment can impact scales beyond the virial radius and with the larger scale environment. 
For instance, redder quiescent galaxies preferentially populate overdense environments \citep[e.g.,][]{dresslerGalaxyMorphologyRich1980, Hogg_2003, Blanton_2005}. 
In this way, conformity has also been observed at much larger (2-4 Mpc) distances, called the "two-halo" conformity \citep[e.g.,][]{kauffmannReexaminationGalacticConformity2013,Olsen_2021,Olsen_2023} with different mechanisms proposed to explain the phenomena \citep[e.g.,][]{Hearin_2015,hearinPhysicalOriginGalactic2016, lacerna_2018, lacernaEnvironmentalInfluenceGroups2022, ayromlouPhysicalOriginGalactic2022,Ayromlou_2023}.

Despite these advances, quantifying how environmental influences vary across different galaxy properties and types remains challenging. Environmental parameters can be defined using various approaches, including spherical overdensities within fixed radii \citep[e.g.,][]{Blanton_2006,Tinker_2008}, distances to $n$th-nearest neighbors \citep[e.g.,][]{Peng_2010,Woo_2013,Banerjee_2021}, or topological features of the cosmic web \citep[e.g.,][]{Sousbie_2011,Kono_2020,Galarraga_2022,Hasan_2024,Wang_2024}. 
Recent work by \citet{Wu_2024} identified optimal length scales for quantifying environmental effects using ML, finding 3 Mpc scales as the most informative for spherical overdensity metrics. In this work, we go beyond the optimal environmental scale for predicting galaxy properties, to answer the following: \textit{"How does the influence of environment on galaxy properties vary across different galaxy populations?"}, and \textit{"Which environmental parameters most strongly influence different galaxy populations?"}

To address these questions, we develop neural network models that predict the $M_*$ and the SFR from subhalo properties of both the host subhalo and surrounding subhalos in the IllustrisTNG simulation. We then apply SHapley Additive exPlanations (SHAP) to quantitatively assess environmental influences and analyze environment-galaxy property correlations. 

This work builds upon various successful demonstrations of ML methods trained on hydrodynamic simulations in predicting baryonic properties from DM subhalos \citep[e.g.,][]{Kamdar_2016,Agarwal_2018, Zhang_2019, Chuang_2024,Wu_2024}.
In contrast to these previous studies, our research explicitly defines the physical properties of individual neighboring galaxies as environmental inputs rather than using density metrics or distances to structural features, and employs SHAP analysis to quantitatively interpret their contributions. 
This interpretable ML approach provides insights into the underlying physical mechanisms driving environment-galaxy relationships.

The paper is structured as follows. In Section~\ref{sec:methods}, we describe simulation data and ML models. Section~\ref{sec:results} presents prediction performance and SHAP analysis. Section~\ref{sec:discuss} interprets the models and discusses environmental influences. Section~\ref{sec:conclusions} summarizes our findings. Several detailed results and discussions can be found in the Appendices. The IllustrisTNG simulation used in this work assumes a Planck 2016 cosmology \citep{Plank_2016}.

\section{Methods} \label{sec:methods}
\subsection{Simulation Data}
We use the outputs of the IllustrisTNG cosmological magnetohydrodynamic simulation \citep{TNG2, TNG1, TNG3, TNG4, TNG5}, which traces the evolution of DM, gas, stars, and black holes from redshift $z = 127$ to $z = 0$. We use the TNG300-1 simulation run, which has the highest mass resolution run among the largest volume, 300 Mpc in the IllustrisTNG suite. This simulation run enables both statistically significant sampling of large-scale structures and the study of processes down to galactic scales. 

\begin{figure}
    \centering
    \includegraphics[width=0.75\linewidth]{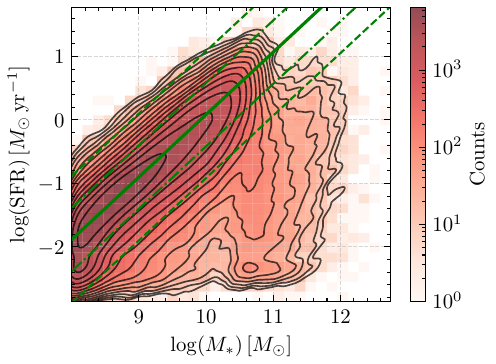}
    \caption{All datasets used in this study are shown. To define the star-formation main sequence (MS; green solid line), kernel density estimation (KDE) is applied to the SFR–$M_*$ plane of the entire dataset, and principal component analysis (PCA) is performed on the top 1\% highest-density region. 
    The green dashed and dash-dotted lines indicate deviations of $\pm1$ dex and $\pm0.5$ dex from the MS, respectively.
    }
    \label{Fig:MS}
\end{figure}

\begin{figure}
    \centering
    \includegraphics[width=1.0\linewidth]{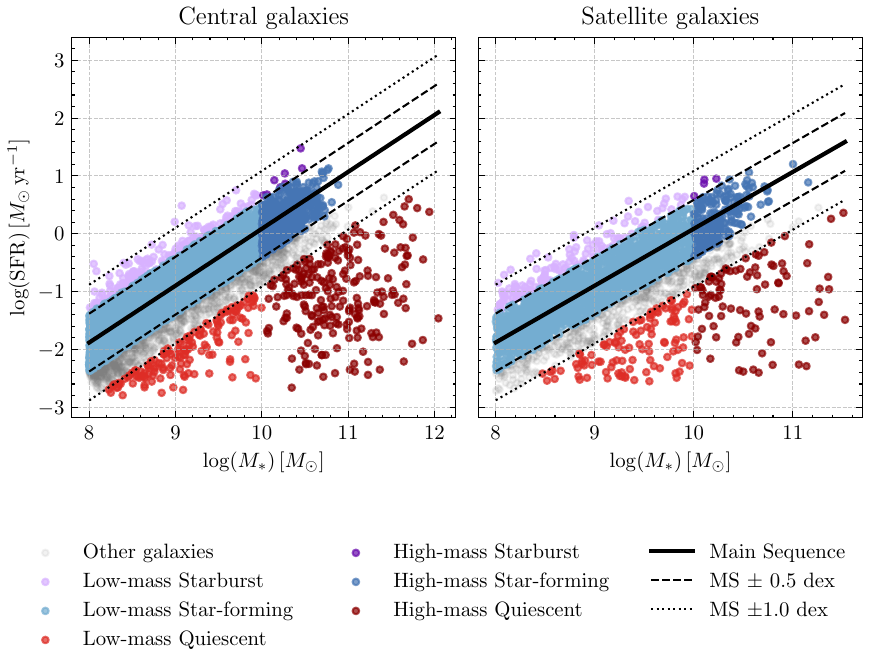}
    \caption{Test set example. Main sequence (MS) derived by Fig.~\ref{Fig:MS}. The colors of the plot are indicative of different galaxy types. 
    }
    \label{fig:6groups}
\end{figure}

We utilize the \texttt{SUBFIND} subhalo catalog at $z=0$, removing flagged subhalos (\texttt{SubhaloFlag != 0}) and selecting galaxies that satisfy $\log({M_h/M_{\odot}})>10$, $\log({M_*/M_{\odot})}>8$, and $\mathrm{SFR}/(M_{\odot}{\mathrm{yr}^{-1}})>0$ (Fig.~\ref{Fig:MS}, and see also \ref{sec:data_app}).
From this selected sample of 393,731 galaxies, we randomly select 300,000 for training, 10,000 for validation, and 10,000 for testing (Fig.~\ref{fig:6groups}),
where these three different data sets are mutually exclusive.
In our dataset, we define galaxies as the satellite galaxies if they have a parent galaxy in the subhalo catalog, and otherwise as central galaxies. According to this definition, 74\% of galaxies are central galaxies.

Our models predict either $M_*$ or SFR using two categories of input features: one is properties of the target subhalo itself (Table~\ref{tab:input_target}), and the other is properties of surrounding subhalos (Table~\ref{tab:input_neighbors}). The surrounding environment is defined using the $N$ nearest neighbor galaxies, where $N$ ranges from 0 to 30. For $N=0$, no environmental information is included, while for $N>0$, the properties of the $N$ closest galaxies are incorporated.

\begin{table}
    \centering
    \caption{Halo properties used as input features for the target galaxy.}
    \begin{tabular}{p{1.9cm} p{5.8cm}} \hline \hline
        notation & Feature  \\ \hline
        $(x,y,z)$ & 3D comoving coordinates \\
        $(v_x, v_y, v_z)$ & Peculiar velocity components\\
        $M_h$ & Subhalo mass \\
        $r_{50}$ & The half $M_h$ radius \\
        $V_{\mathrm{max}}$ & The maximum circular velocity of halo \\ \hline
    \end{tabular}
    \label{tab:input_target}
\end{table}

\begin{table}
    \centering
    \caption{Halo properties used as input features for the surrounding galaxies.}
    \begin{tabular}{p{1.9cm} p{5.8cm}} \hline \hline
        notation & Feature  \\ \hline
        $(D, \theta, \phi)$ & Relative coordinates with respect to the target galaxy\\
        $(v_D, v_{\theta}, v_{\phi})$ & Relative peculiar velocity components\\
        $M_h$ & Subhalo mass \\
        $r_{50}$ & The half $M_h$ radius \\
        $V_{\mathrm{max}}$ & The maximum circular velocity of halo \\ \hline
    \end{tabular}
    \label{tab:input_neighbors}
\end{table}

\begin{figure}
    \centering
    \includegraphics[width=0.7\linewidth]{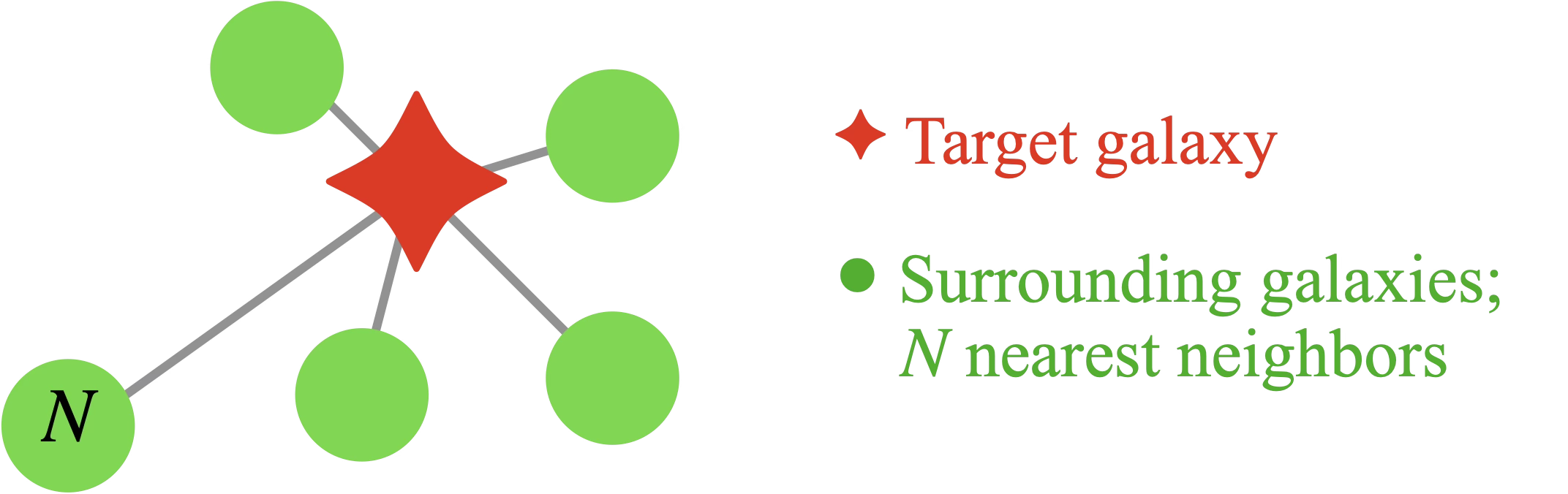}
    \caption{Schematic illustration of the definition of the surrounding environment. 
    The parameter $N$ takes values $0 \sim 10$, $15$, $20$, $25$, and $30$, where $N = 0$ indicates the absence of any environmental information.
    }
    \label{Fig:target-surround}
\end{figure}

\subsection{Neural Network Model}
We employ a Multi-Layer Perceptron (MLP) with seven hidden layers of sizes [128, 226, 226, 512, 128, 128, 16]. The number of inputs is $9\times(1+N)$, depending on the number of neighbors included, and the output is a single scalar value. The activation function is Rectified Linear Unit (ReLU), and we use mean squared error (MSE) as the loss function.
\begin{align}
    \text{MSE}& = \frac{1}{n} \sum_{i=1}^{n} (y_{\text{true}, i} - y_{\text{pred}, i})^2 \\
    \text{where:} \nonumber \\
    y_{\text{true}, i} &:\ \text{the actual value for sample } i \nonumber \\
    y_{\text{pred}, i} &:\ \text{the predicted value for sample } \nonumber i \\
    \bar{y}_{\text{true}} &:\ \text{the mean of the actual values} \nonumber \\
    n &:\ \text{the number of samples} \nonumber
\end{align}
The model is optimized using AdamW \citep{loshchilov2019decoupledweightdecayregularization} with learning rate and weight decay both set to $10^{-4}$. Layer Normalization is applied to each hidden layer, with dropout (probability 0.5) on all hidden layers except the final one. Batch sizes are 64, 16, and 16 for training, validation, and testing, respectively.
We adopt early stopping, terminating the model training when the validation error fails to decrease for 10 consecutive epochs.
All models are implemented in PyTorch \citep{pytorch}.

To account for model performance variability, we train 10 models with different random data splits.
For each model, we compute the $R^2$, the coefficient of determination,
\begin{align}
    R^2 &= 1 - \frac{\sum_{i=1}^{n} (y_{\text{true}, i} - y_{\text{pred}, i})^2}
                 {\sum_{i=1}^{n} (y_{\text{true}, i} - \bar{y}_{\text{true}})^2}
\end{align}
and select the model and data splits which provide a median $R^2$ value.
Once the model is fixed, we then generate 100 bootstrap samples from the test set by random sampling with replacement, where the sample size is 100\% of the parent sample.
The distribution of the $R^2$ values is used to robustly compare the models with different $N$ neighbors.

\subsection{Explainable AI} \label{sec: explain_shap}
To interpret our models, we use SHapley Additive exPlanations \citep[SHAP:][]{SHAP_NIPS2017_7062}, 
a game-theoretic approach that quantifies each feature’s contribution to a model’s prediction.
SHAP assigns a contribution value (SHAP value) to each feature for an individual prediction, such as the predicted $M_*$ or SFR of each galaxy in the dataset.

SHAP analysis is increasingly used for interpreting neural networks in astrophysical contexts \citep[e.g.,][]{2022MNRAS.512.1710H, iwasakiExtractingInformativeLatent2023, alfonzoKatachiXingDecoding2024, Wu_2024, 2025MNRAS.537..876A}. 
Positive SHAP values indicate features that increase predictions relative to a baseline, while negative values indicate features that decrease predictions. 
The sum of SHAP values equals the difference between the prediction and the baseline. 
SHAP values are computed for each galaxy, enabling us to analyze feature contributions individually. 
For example, we can assess which environmental or intrinsic features most influence $M_*$ or SFR for a given galaxy. 
For a more detailed explanation of SHAP, see \citet{molnarInterpretableMachineLearning}. 

We use SHAP in this study to understand how predictions of $M_*$ and SFR are made as a function of the input environmental features. 
We calculate SHAP values using the mean prediction from 10,000 randomly selected training samples as our baseline.
This allows us to quantitatively assess which environmental and intrinsic features most strongly influence $M_*$ and SFR predictions for different galaxy populations.
We denote the SHAP value as $S_i^j$ for $i$-th input for $j$-th galaxy.
    
\section{Results}\label{sec:results}
\subsection{Model output} \label{sec:result_model}
\begin{figure*}
    \centering
    \begin{subfigure}{0.95\textwidth}
        \centering
        \includegraphics[width=\linewidth]{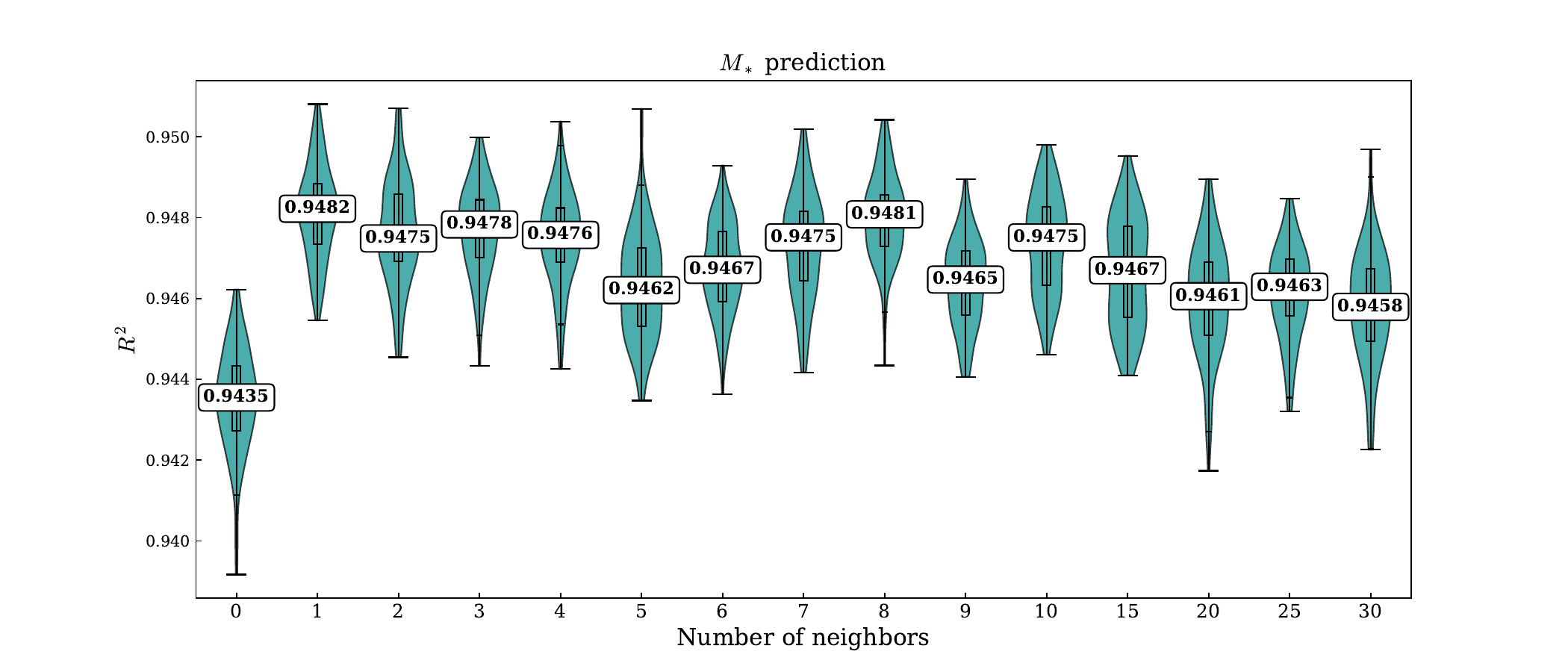}
    \end{subfigure}
    \hfill
    \begin{subfigure}{0.95\textwidth}
        \centering
        \includegraphics[width=\linewidth]{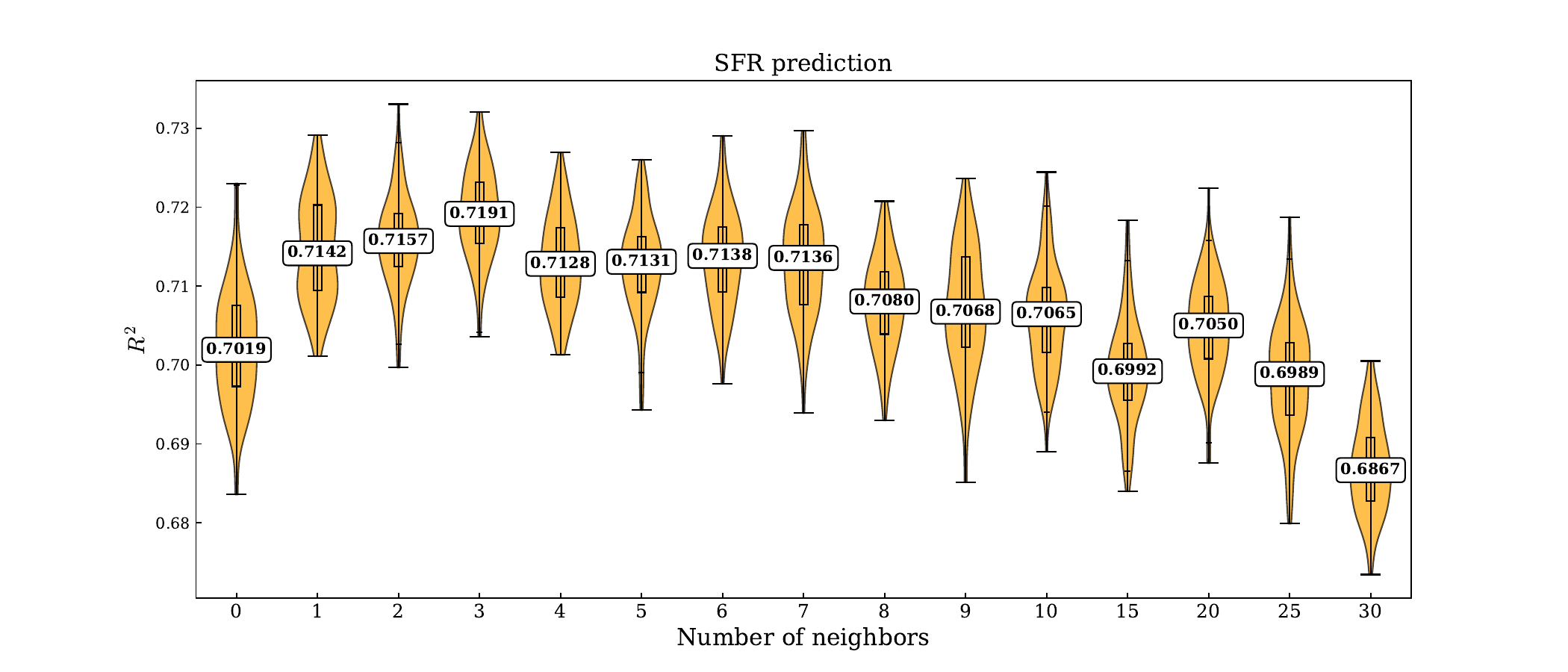}
    \end{subfigure}
    \caption{Distribution of coefficient of determination ($R^2$) values for models with varying numbers of neighboring galaxies. Each distribution is generated through 100 bootstrap tests, with median values displayed above each violin plot. The one-neighbor model demonstrates the highest median $R^2$ for $M_*$ predictions (top), while the three-neighbors model yields the highest median $R^2$ for SFR predictions (bottom).}
    \label{Fig.violin}
\end{figure*}

We perform 100 times bootstrap on the test set for each given model to compute the distribution of $R^2$. Then we compare the distribution of $R^2$ among models with different numbers of neighbors as shown in Fig.~\ref{Fig.violin}.

For $M_*$ prediction, the Kolmogorov–Smirnov test on the $R^2$ distribution reveals that all models incorporating environmental information show statistically significant improvements at the 1\% significance level compared to the zero-neighbor model. 
Among these, the one-neighbor model shows the most substantial improvement in the median $R^2$ value. 
Furthermore, the one-neighbor model exhibits statistically significant differences at the 1\% level when compared with all other models, except for the three- and eight-neighbors models.

For SFR prediction, the one- to nine-neighbors models demonstrate statistically significant improvements over the zero-neighbor model at the 1\% level.
Among these, the three-neighbors model shows the most improvement in the median $R^2$ value, with statistical significance ($p < 0.01$).

These results indicate that incorporating the halo properties of surrounding galaxies helps the prediction accuracy of $M_*$ and SFR. For $M_*$ prediction, considering only the nearest neighboring galaxy provides optimal performance, while for SFR prediction, including information up to the third-nearest galaxy yields the highest accuracy. Based on these findings, we employ the one-neighbor model for $M_*$ prediction and the three-neighbors model for SFR prediction in subsequent analyses.

\begin{figure}
    \centering
    \begin{minipage}{0.235\textwidth}
        \centering
        \includegraphics[width=\linewidth]{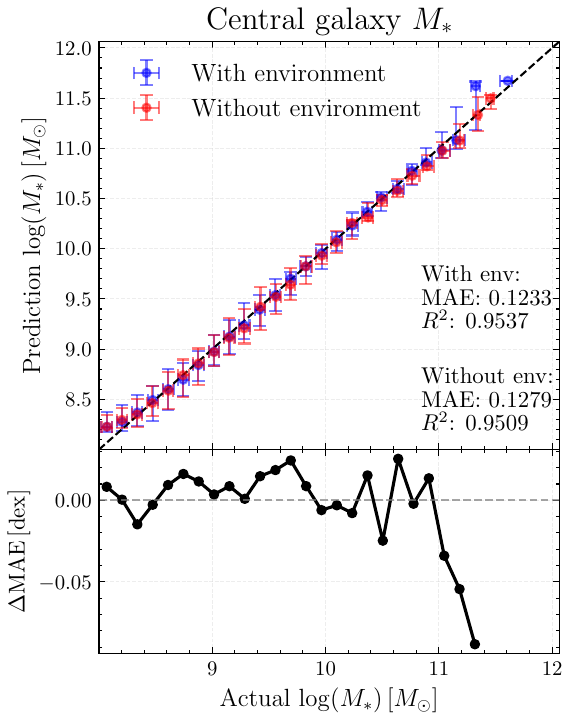}
    \end{minipage}
    \begin{minipage}{0.235\textwidth}
        \centering
        \includegraphics[width=\linewidth]{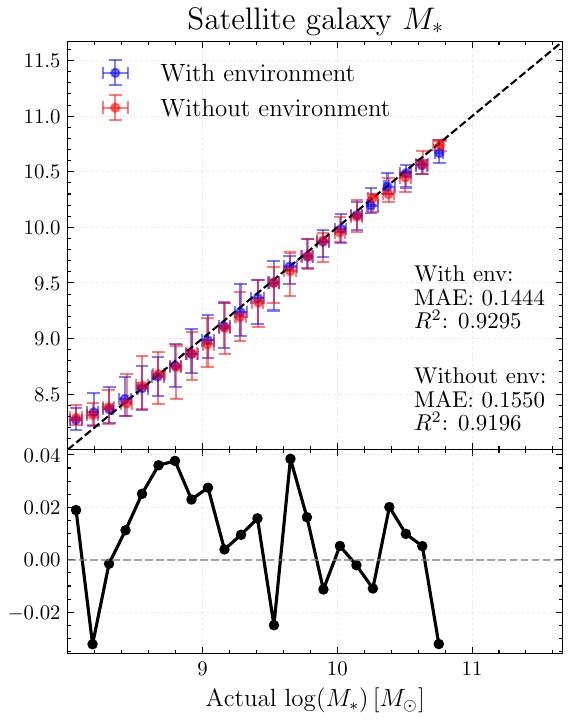}
    \end{minipage}
    \caption{
    Comparison of $M_*$ predictions for central (left) and satellite galaxies (right) with and without environmental information. 
    Top panel: Blue and red points represent median values in each mass bin for models with (one-neighbor model) and without environmental information, respectively. Error bars indicate 16th and 84th percentile ranges. Dashed lines indicate perfect one-to-one relationships.  
    Bottom panel: Difference in mean absolute error (MAE), i.e., difference in model prediction accuracy,
    between predictions without and with environmental information as a function of $M_*$ (calculated as $\mathrm{MAE}_{\mathrm{without}} - \mathrm{MAE}_{\mathrm{with}}$ at each bin). Positive values indicate improved performance with environmental information. Horizontal dashed lines mark $\Delta \mathrm{MAE} = 0$.
    Overall MAE and $R^2$ values for each model are displayed in top panels. Each bin contains at least 10 galaxies to ensure statistical reliability.}
    \label{Fig.pre_sm}
\end{figure}

\begin{figure}
    \centering
    \begin{minipage}{0.235\textwidth}
        \centering
        \includegraphics[width=\linewidth]{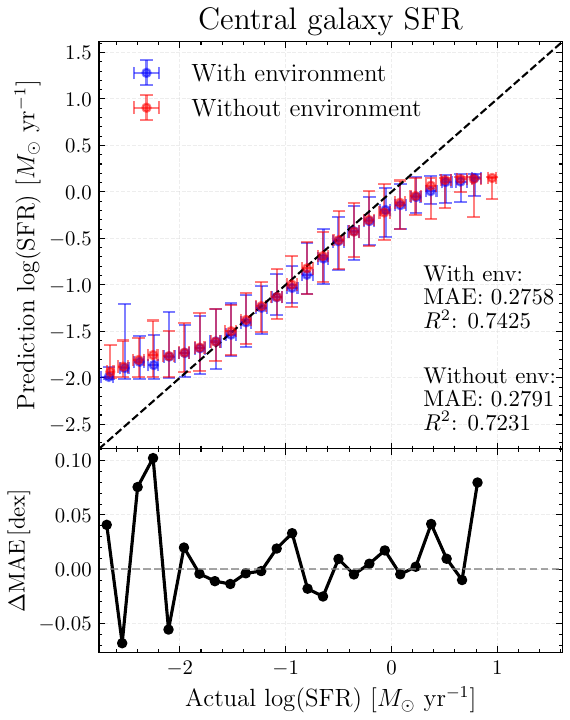}
    \end{minipage}
    \begin{minipage}{0.235\textwidth}
        \centering
        \includegraphics[width=\linewidth]{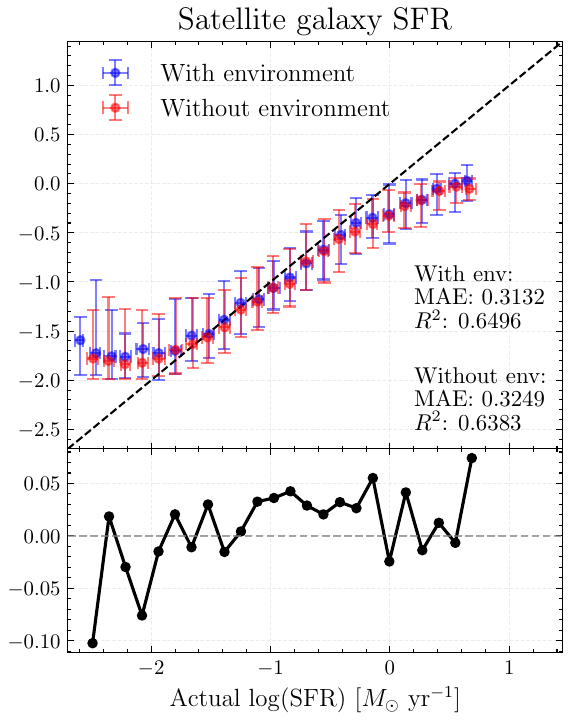}
    \end{minipage}
    \caption{
    Comparison of SFR predictions for central (left) and satellite galaxies (right) with and without environmental information. The panel layout and formatting follow the conventions established in Fig.~\ref{Fig.pre_sm}. The "with environment" case corresponds to results from the three-neighbors model. SFR values are predicted based on the target galaxy's halo properties and those of its three nearest neighbors. 
    }
    \label{Fig.pre_sfr}
\end{figure}

Fig.~\ref{Fig.pre_sm} and~\ref{Fig.pre_sfr} present the predictions of $M_*$ and SFR, respectively, for central and satellite galaxies with and without environmental information. As shown in the bottom panels of these figures, the prediction accuracy for $M_*$ improves significantly at lower masses ($\log(M_*/M_{\odot}) < 10$), whereas $M_*$ of high-mass galaxies can be accurately predicted solely from their intrinsic properties, with no additional benefit from environmental information.
For SFR predictions, we observe no significant improvement for central galaxies, while for satellite galaxies, the prediction accuracy increases for actively star-forming galaxies ($\log(\mathrm{SFR}/ (M_{\odot}{\mathrm{yr}^{-1}})) > -1$). For further details on the differences between models with and without environmental information, see \ref{sec:appe_diffe_with_without}.

\begin{figure}
    \centering
    \begin{minipage}{0.235\textwidth}
        \centering
        \includegraphics[width=\linewidth]{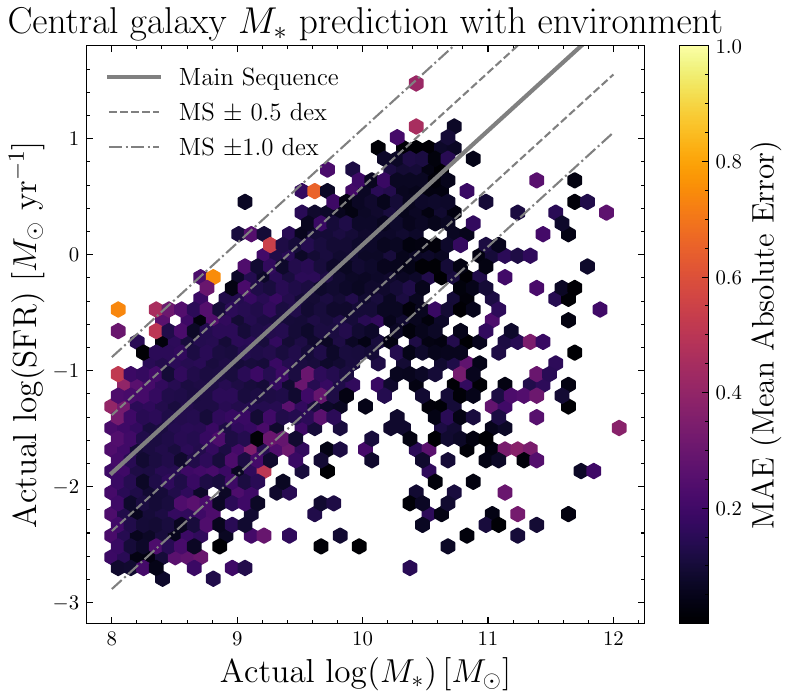}
    \end{minipage}
    \begin{minipage}{0.235\textwidth}
        \centering
        \includegraphics[width=\linewidth]{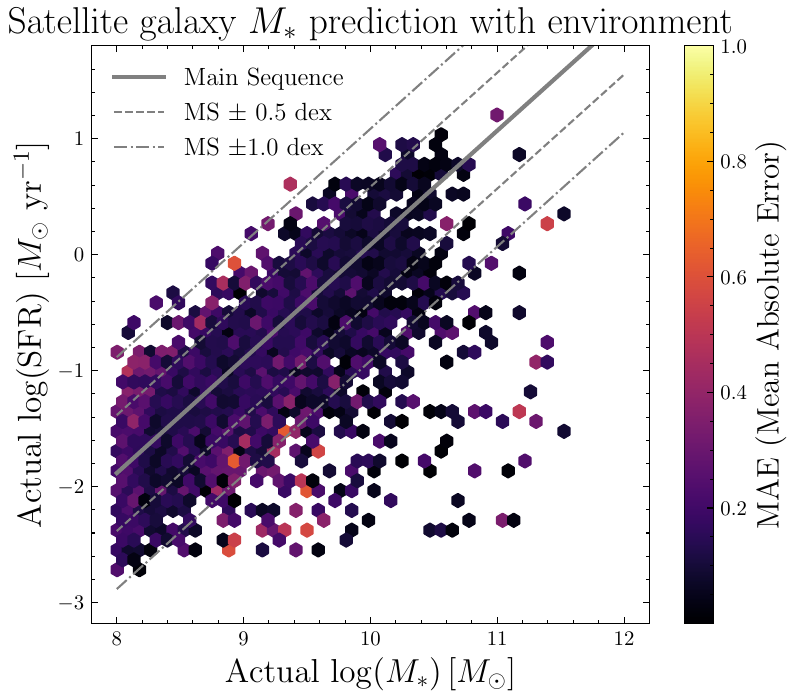}
    \end{minipage}
    \vspace{5mm}
    \begin{minipage}{0.235\textwidth}
        \centering
        \includegraphics[width=\linewidth]{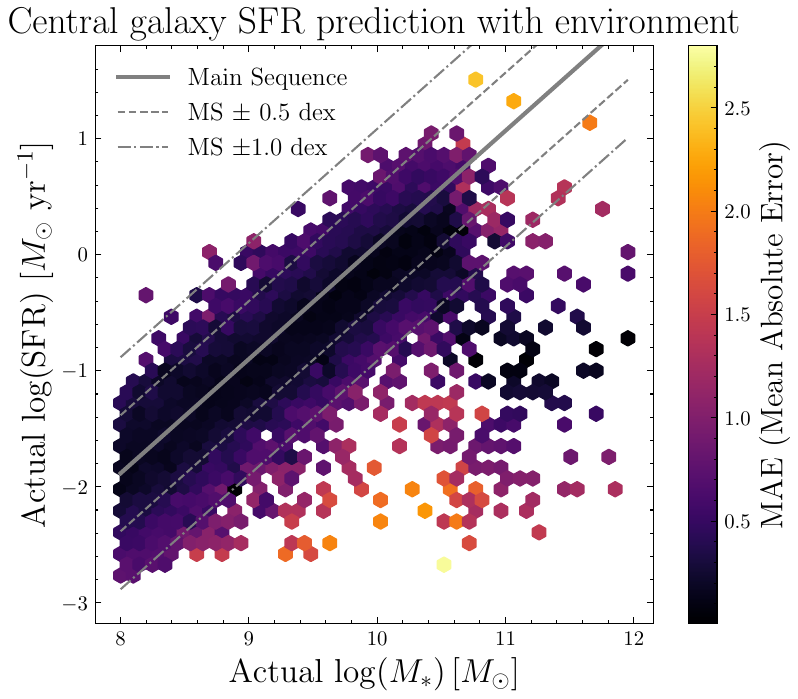}
    \end{minipage}
    \begin{minipage}{0.235\textwidth}
        \centering
        \includegraphics[width=\linewidth]{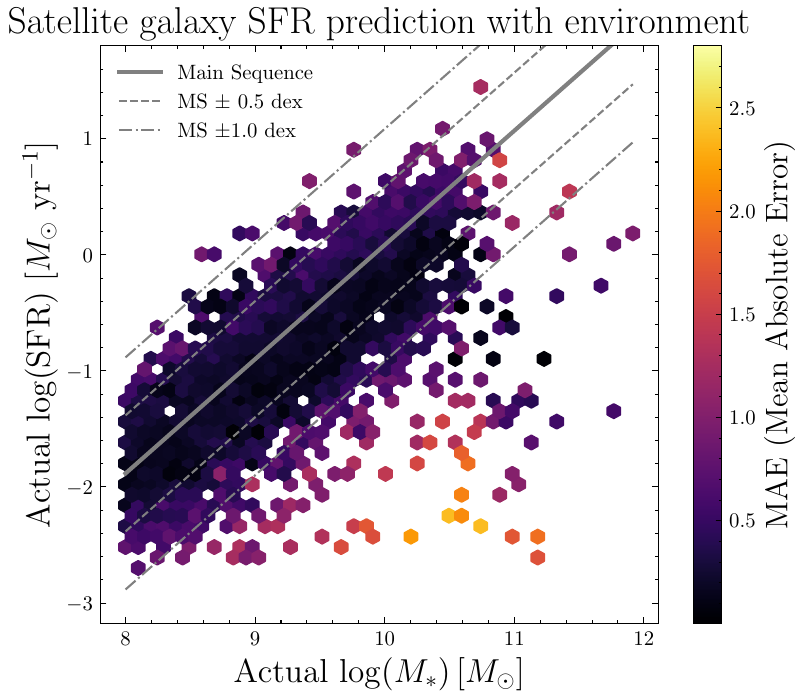}
    \end{minipage}
    \caption{The MAE distribution in the SFR-$M_*$ plane for the $M_*$ predictions (one-neighbor model) and SFR predictions (three-neighbors model) for central and satellite galaxies, corresponding to Fig.~\ref{Fig.pre_sm} and ~\ref{Fig.pre_sfr} results.}
    \label{Fig.MAE_distribu}
\end{figure}

Fig.~\ref{Fig.MAE_distribu} presents the prediction accuracy on the SFR–$M_*$ plane for each test dataset using the mean absolute error (MAE):
\begin{equation}
    \text{MAE} = \frac{1}{n}\sum_{i=1}^{n} \left| y_{\text{true}, i} - y_{\text{pred}, i} \right|
\end{equation}
The Star-Forming Main Sequence (MS) is defined for the entire dataset (Fig.~\ref{Fig:MS}), with dashed lines indicating regions 0.5 dex and 1 dex away from the MS.
Our model exhibits significant variations in prediction accuracy across different galaxy types, particularly for SFR predictions. The accuracy is notably higher for galaxies near the MS, with systematically lower accuracy for low-mass quenched galaxies.

\subsection{SHAP Analysis: Contribution from the environment} \label{sec:result_shap}
In this section, we will see that which input attributes have importance on the final prediction of $M_*$ and SFR. As seen from the analysis in the previous section, the information content originates from different sources, depending on their $M_*$ and SFR. Therefore, here we divide the test sample into six different groups based on $M_*$ and SFR and then apply the SHAP (SHapley Additive exPlanations) analysis to the sample in each group \citep[e.g.,][]{rodighieroLESSERROLESTARBURSTS2011, donnariStarFormationActivity2019a, Schreiber2015, yuanUnravelingEmissionLine2025}.

First, we classify galaxies by stellar mass as: 
(i) Low-mass: $\log(M_*/M_{\odot}) < 10$,
(ii) High-mass: $\log(M_*/M_{\odot}) > 10$.
Then, we further subdivide based on star formation rate as:
(a) Starburst: $\log(\mathrm{SFR}) > \mathrm{MS} + 0.5 \, \mathrm{dex}$, 
(b) Star-forming: $\mathrm{MS} - 0.5 \, \mathrm{dex} < \log(\mathrm{SFR}) < \mathrm{MS} + 0.5 \, \mathrm{dex}$, and
(c) Quiescent: $\log(\mathrm{SFR}) < \mathrm{MS} - 1 \,  \mathrm{dex}$.
By combining these classifications, we establish six galaxy groups as illustrated in Fig.~\ref{fig:6groups}: Low-mass Starburst galaxies, Low-mass Star-forming galaxies, Low-mass Quiescent galaxies, High-mass Starburst galaxies, High-mass Star-forming galaxies, and High-mass Quiescent galaxies. In each group, we further differentiate between central and satellite galaxies.
    
\begin{figure*}
    \centering
    \begin{subfigure}{0.5\textwidth}
        \centering
        \includegraphics[width=\textwidth]{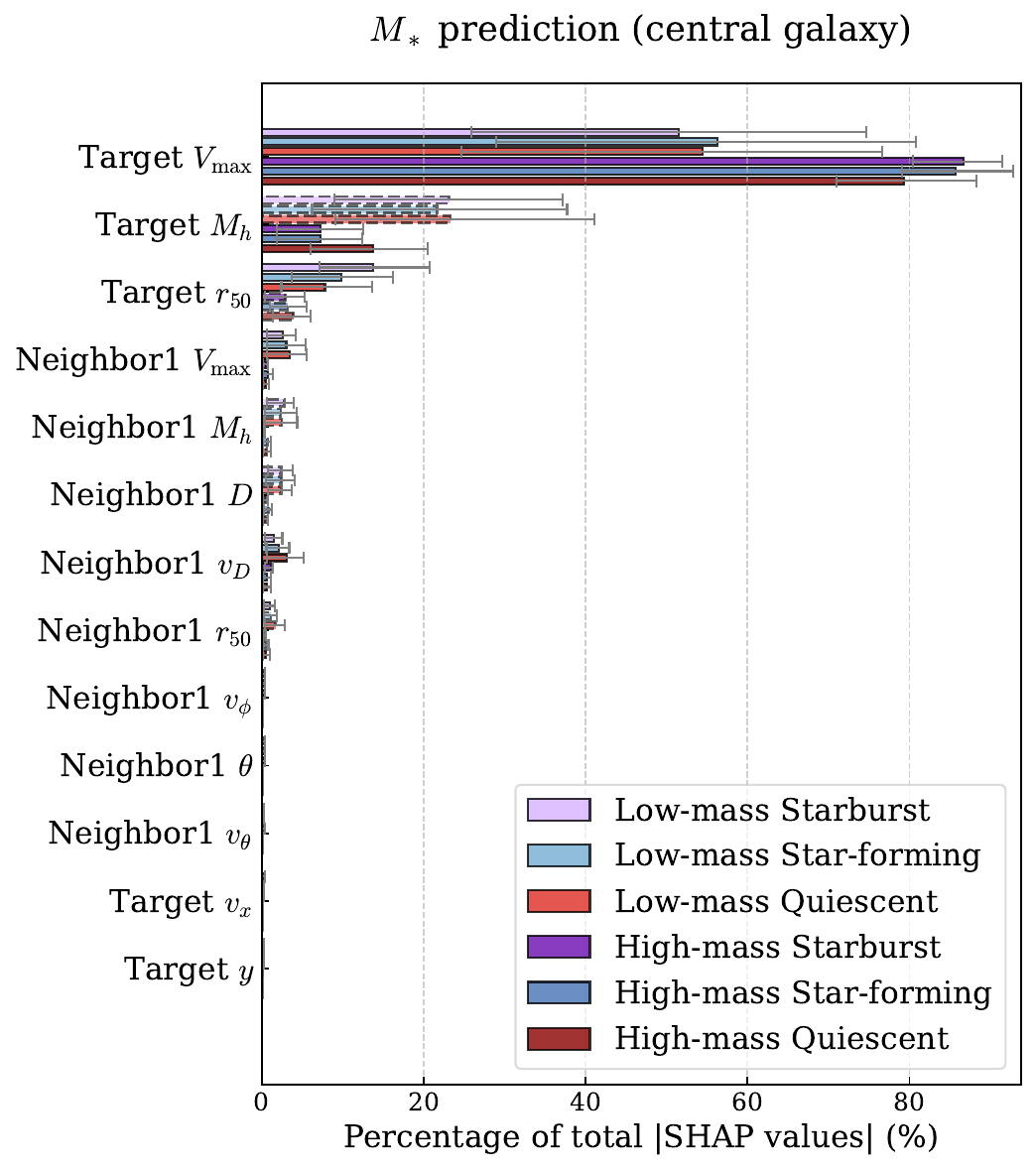}
    \end{subfigure}%
    \hfill
    \begin{subfigure}{0.5\textwidth}
        \centering
        \includegraphics[width=\textwidth]{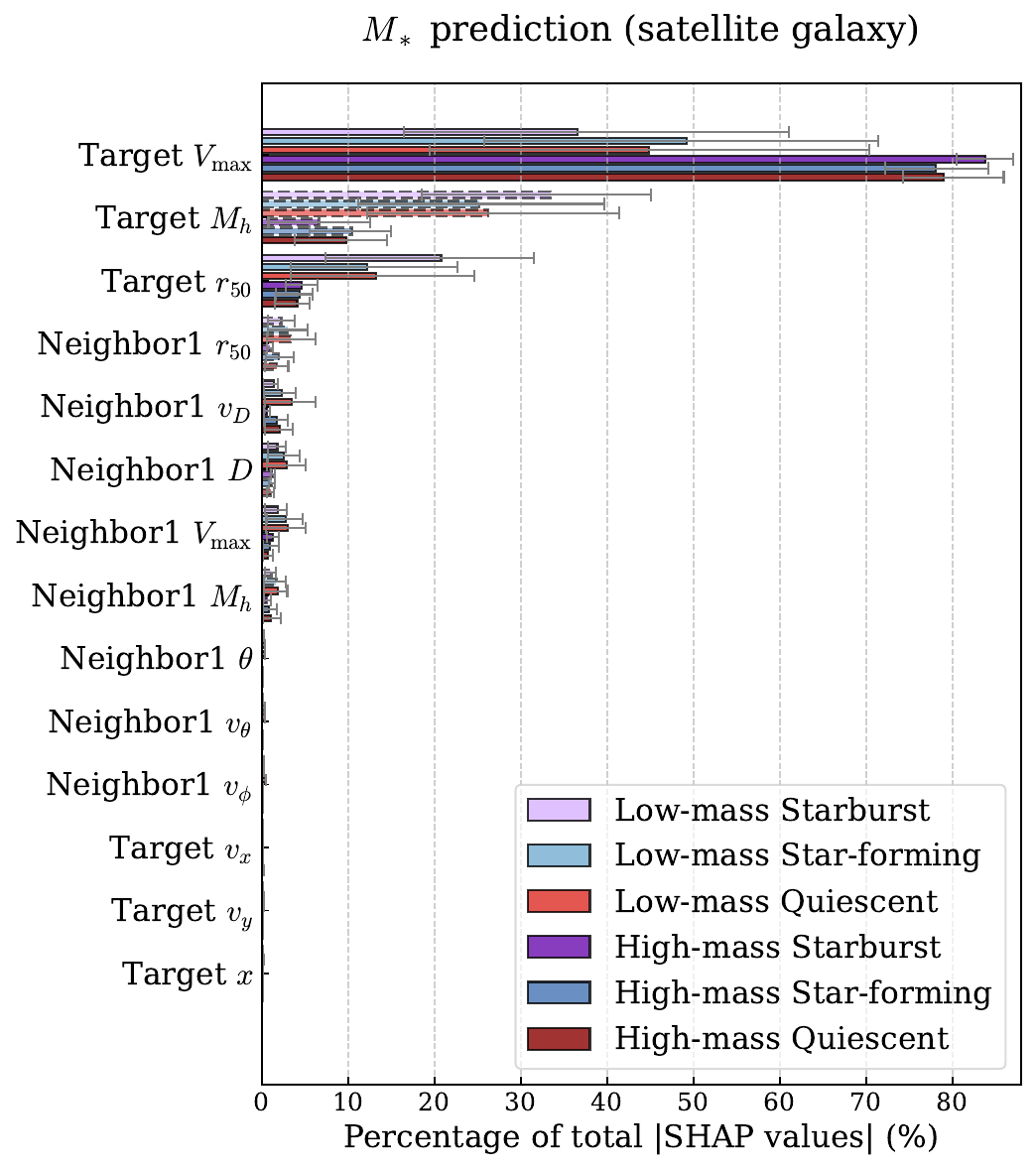}
    \end{subfigure}
    \noindent\hdashrule{\textwidth}{0.5pt}{3pt 3pt}
    \begin{subfigure}{0.5\textwidth}
        \centering
        \includegraphics[width=\textwidth]{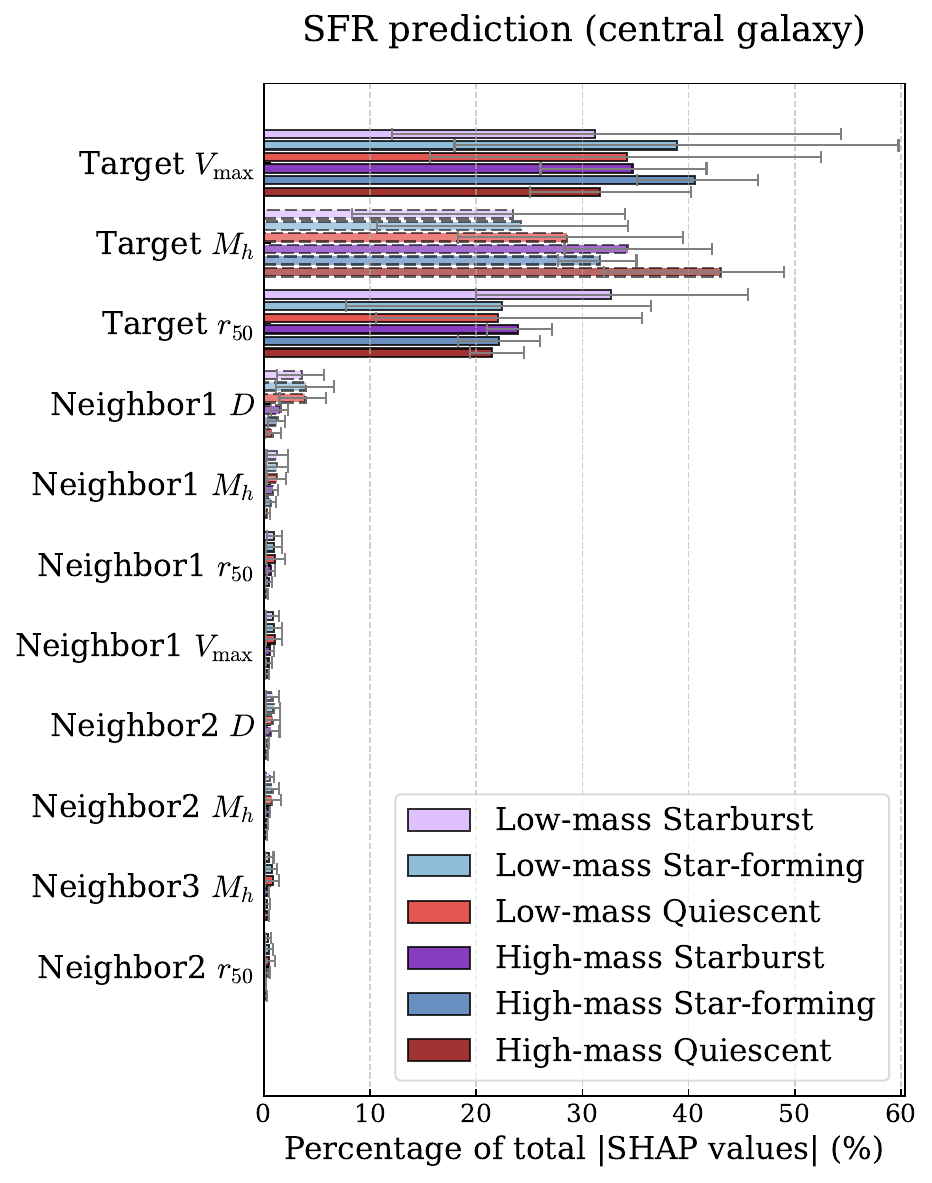}
    \end{subfigure}%
    \hfill
    \begin{subfigure}{0.5\textwidth}
        \centering
        \includegraphics[width=\textwidth]{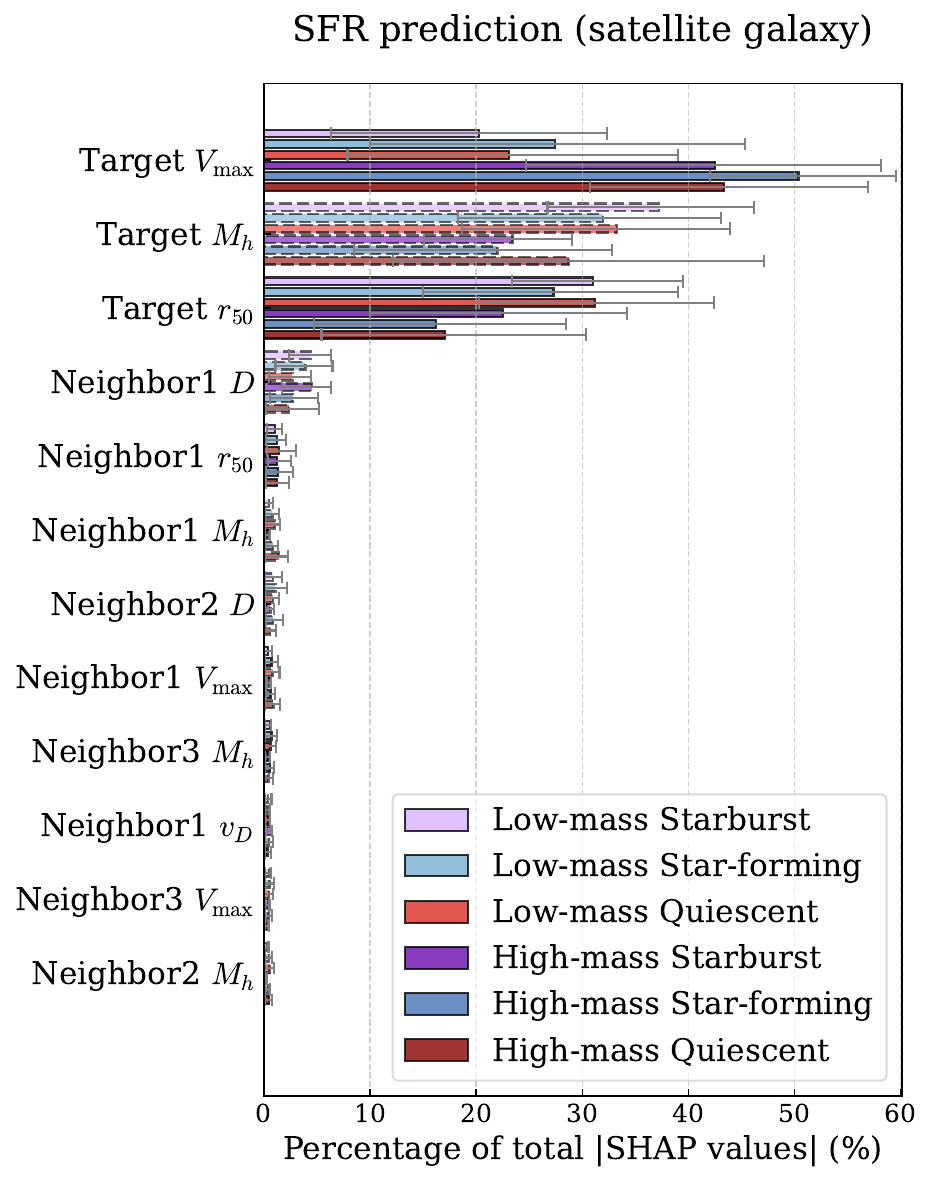}
    \end{subfigure}
    \caption{
        Top ten input features contributing to the prediction of $M_*$ (upper panels) and SFR (lower panels) for central (left) and satellite galaxies (right). 
        The vertical axis represents the union of the top 10 most important features for each galaxy type. 
        In contrast, the horizontal axis shows the contribution ratio of each feature, calculated as the absolute SHAP value relative to the total.
        Each bar represents the mean value, and the error bars indicate the 16th and 84th percentiles.
        Solid and dashed lines around the bars show the mean positive and negative contributions, respectively, associated with increasing feature values.
        Features are ordered by their average contribution across all galaxy types, with the most influential features appearing at the top.
    }
    \label{Fig.SHAP_feature}
\end{figure*}

We compute SHAP values for the best-performing models: the one-neighbor model for $M_*$ prediction and the three-neighbors model for SFR prediction. Fig.~\ref{Fig.SHAP_feature} displays the top ten input features for each group, ranked in descending order of mean absolute SHAP values.
To indicate the average effect of changes in feature values on the predicted values, the direction of each feature’s contribution is determined by calculating the correlation coefficient between the feature values and their corresponding SHAP values within each sample. 
A positive correlation indicates that an increase in the feature value tends to increase the predicted value relative to the baseline, whereas a negative correlation suggests that an increase in the feature value tends to decrease the predicted value relative to the baseline.
For both $M_*$ and SFR, the significant fraction of predictive information derives from intrinsic DM halo properties, particularly $V_{\mathrm{max}}$, $M_h$, and $r_{50}$.
In contrast, the position and velocity of the target galaxy, along with the relative coordinates and velocities of neighboring galaxies in the directions of the declination angles ($\theta$, $\phi$), contribute negligibly ($\sim0.1\%$).

To isolate the environmental contribution, we train new models that exclude parameters related to the target halo's formation history, specifically $V_{\mathrm{max}}$ and $r_{50}$, while retaining $M_h$. The prediction accuracy of these modified models is presented in \ref{sec:appe_predic_no-assembly}.
We quantify the environmental influence as the ratio of SHAP values $S$ summed over all inputs related to the environment to all inputs:
\begin{equation}
    r = \frac{\sum_j\sum_{i\in \text{neighbours input}}|S_i^j|}
    {\sum_j \sum_{i\in \text{all input}} |S_i^j|},
\end{equation}
where the sum index $j$ is taken over all test subsamples for each galaxy population group.

\begin{figure*}
    \centering
    \includegraphics[width=0.85\linewidth]{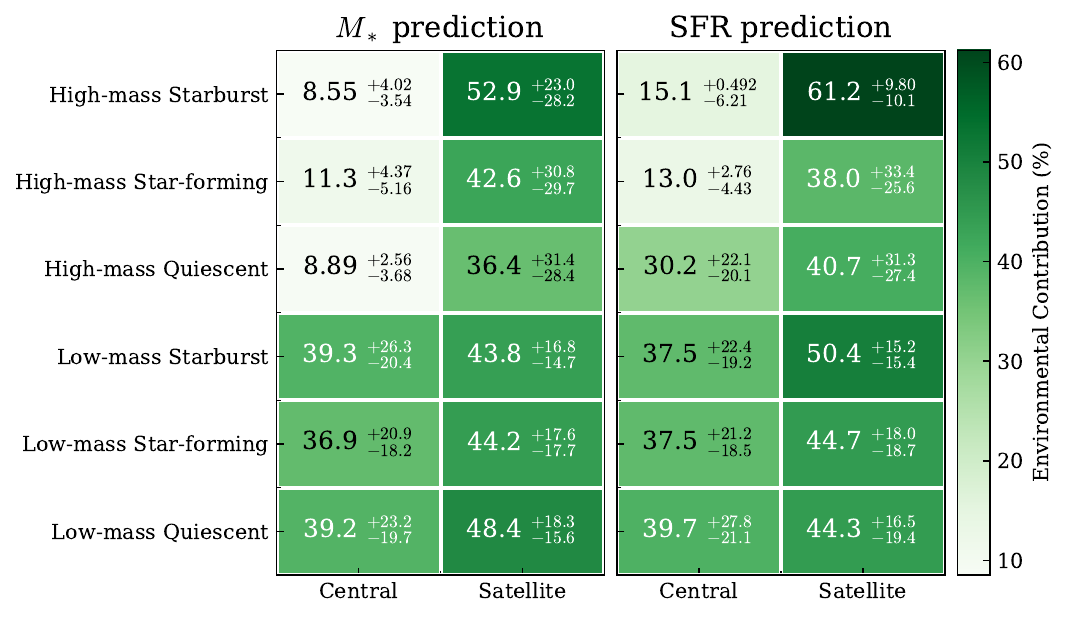}
    \caption{
    Quantification of environmental contributions to $M_*$ and SFR predictions. This analysis is based on models that exclude formation history indicators ($V_{\mathrm{max}}$ and $r_{50}$) and utilize only positional, velocity, and mass information from target and surrounding halos. Values represent the fractional contribution of environmental features to predictions, calculated as the ratio of environmental feature SHAP values to total SHAP values. Points indicate mean values; error bars show 16th and 84th percentiles. The galaxy classification scheme follows Fig.~\ref{fig:6groups}, with results presented separately for the one-neighbor model ($M_*$ prediction) and three-neighbors model (SFR prediction).
    }
    \label{tab:shap_frac_no}
\end{figure*}

Fig.~\ref{tab:shap_frac_no} shows the SHAP fraction $r$ for different galaxy populations, revealing several important patterns:
\begin{enumerate}
    \item Central galaxies experience less environmental influence on both $M_*$ and SFR compared to satellite galaxies.
    \item Among central galaxies, higher-mass systems show significantly reduced environmental dependence.
    \item Among satellite galaxies, higher-mass systems also show a trend toward reduced environmental dependence.
    However, satellite starburst galaxies represent a notable exception, where high-mass systems show a slightly higher dependence on the environment compared to the low-mass systems in both $M_*$ and SFR predictions.
    \item For a given mass bin, no marked difference in the environmental contribution is observed across different star formation activities.

    \end{enumerate}
  
\begin{figure*}
    \centering
    \begin{subfigure}{0.5\textwidth}
        \centering
        \includegraphics[width=\textwidth]{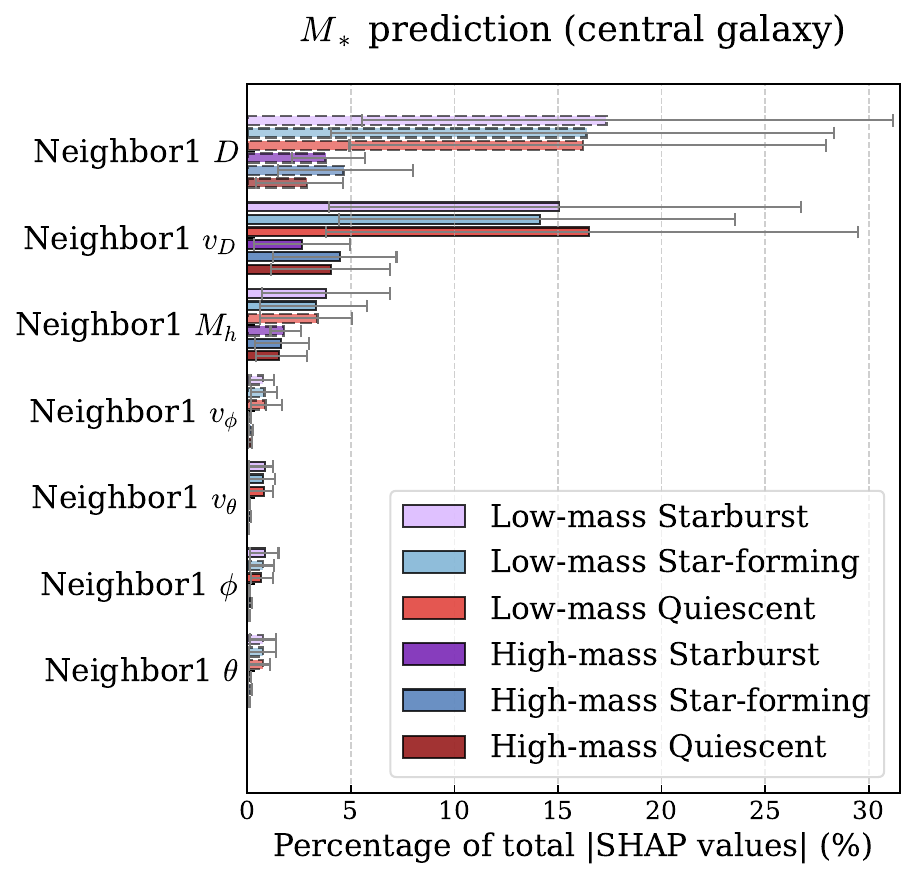}
    \end{subfigure}%
    \hfill
    \begin{subfigure}{0.5\textwidth}
        \centering
        \includegraphics[width=\textwidth]{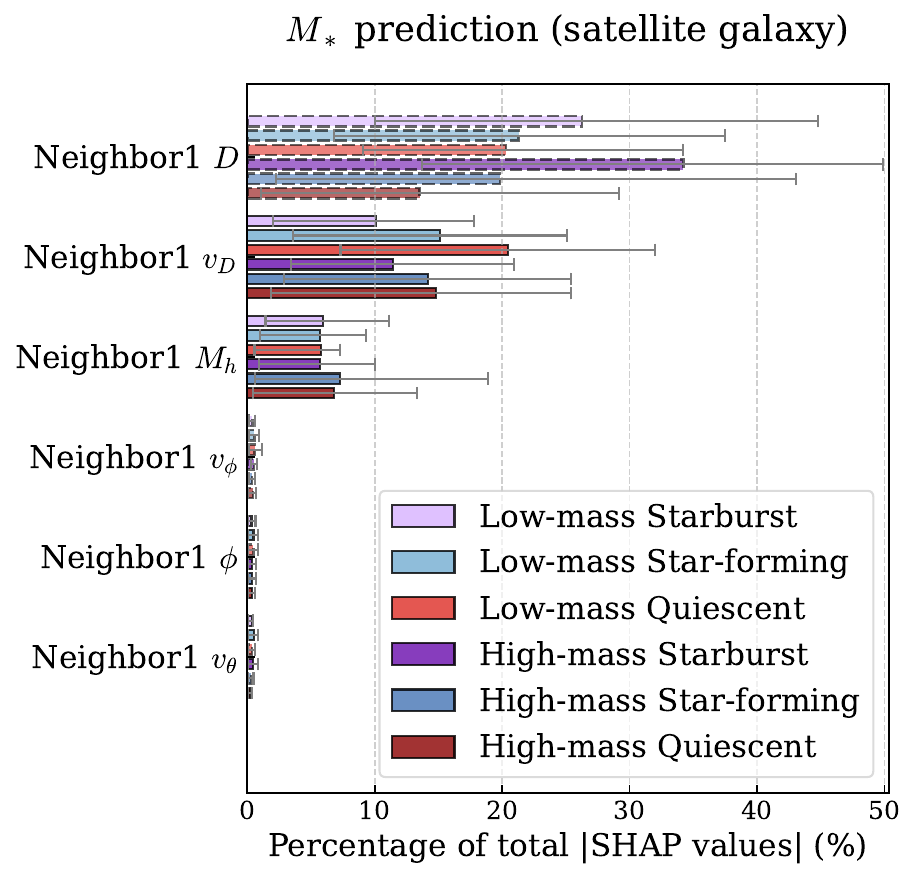}
    \end{subfigure}
    \noindent\hdashrule{\textwidth}{0.5pt}{3pt 3pt}
    \begin{subfigure}{0.5\textwidth}
        \centering
        \includegraphics[width=\textwidth]{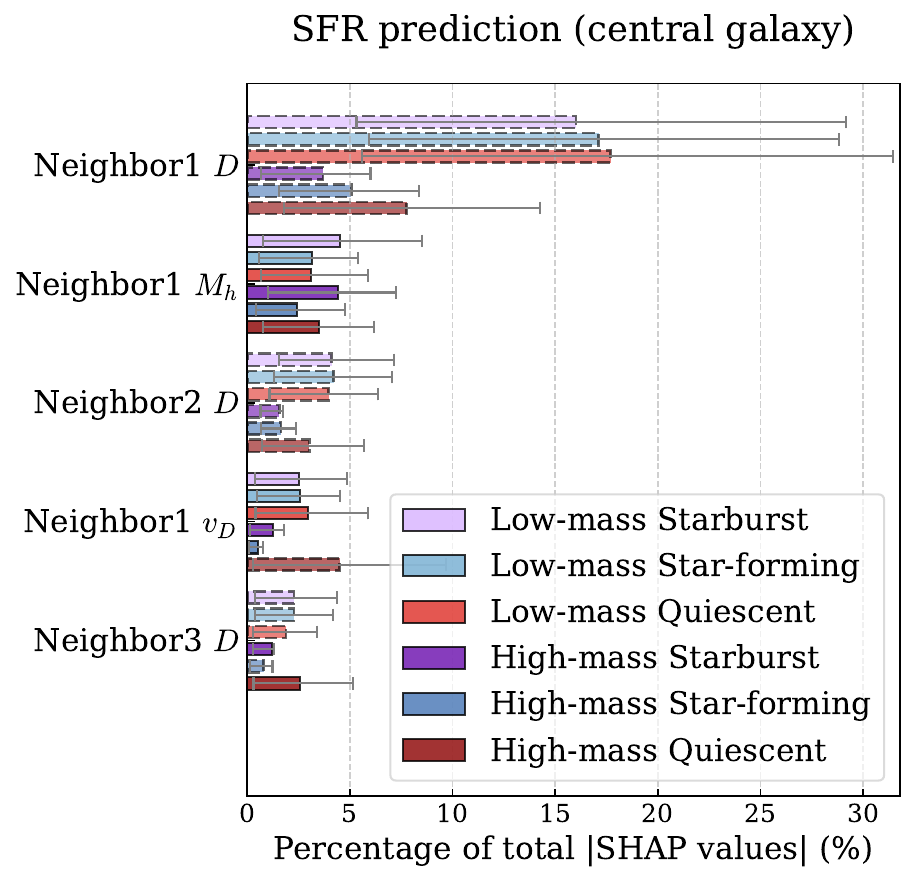}
    \end{subfigure}%
    \hfill
    \begin{subfigure}{0.5\textwidth}
        \centering
        \includegraphics[width=\textwidth]{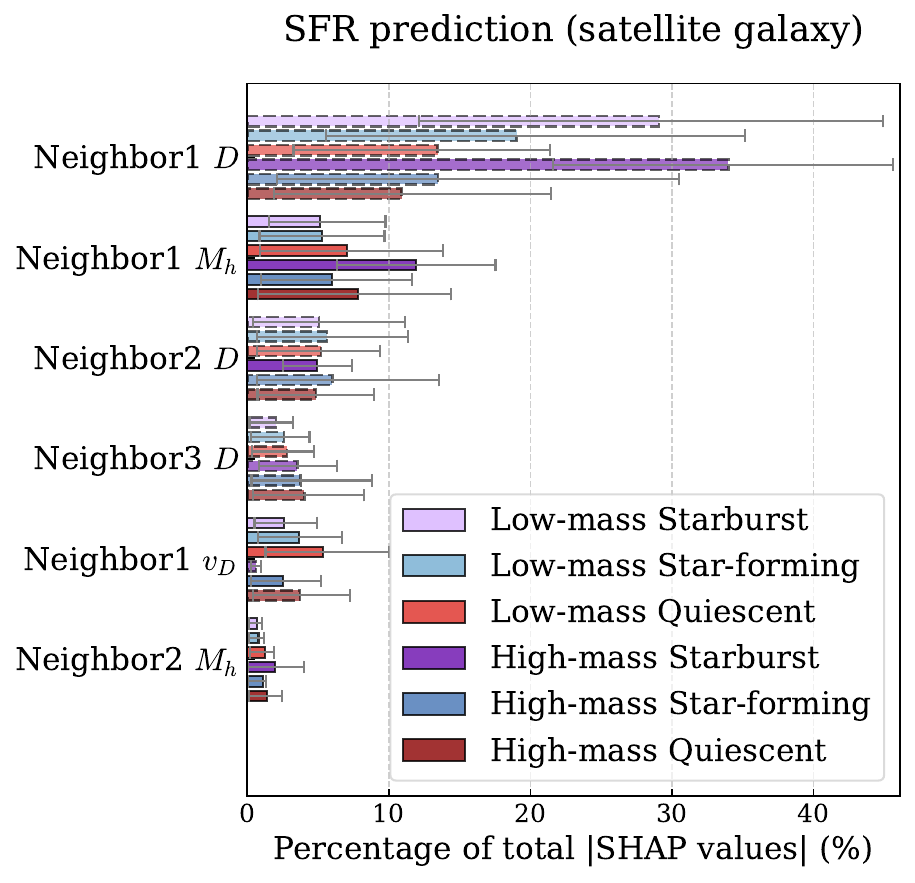}
    \end{subfigure}
    \caption{
    The five most influential environmental features for predicting $M_*$ (upper panels) and SFR (lower panels) in each galaxy population.
    Solid and dashed lines around the bars show the mean positive and negative contributions, respectively, associated with increasing feature values.
    The vertical axis represents the union of top environmental features across galaxy types, with the same horizontal axis definitions as in Fig.~\ref{Fig.SHAP_feature}. 
    This analysis utilizes models that exclude $V_{\mathrm{max}}$ and $r_{50}$ to isolate pure environmental effects, providing a detailed breakdown of the contribution fractions presented in Fig.~\ref{tab:shap_frac_no}. 
    Comprehensive SHAP value results are provided in Fig.~\ref{Fig.SHAP_summary_SM_no} and \ref{Fig.SHAP_summary_SFR_no}.
    }
    \label{Fig.SHAP_feature_no}
\end{figure*}

Fig.~\ref{Fig.SHAP_feature_no} presents the five most significant input environmental features for each group, ranked in descending order of their mean absolute SHAP values when including $M_h$. 
For the $M_*$ prediction, the distance to the nearest galaxy, $D$ and the relative peculiar velocity, $v_D$ have the highest average contributions among the environmental features.
Similarly, for the SFR prediction, the nearest galaxy’s $D$ and $v_D$, along with the $M_h$, exhibit significant contributions among the environmental features.
Notably, $D$ tends to have a greater mean contribution than neighbor $M_h$.
Furthermore, it is evident that as the value of the nearest galaxy's $D$ increases, both $M_*$ and SFR also tend to decrease on average.

\begin{figure}
    \centering
    \begin{minipage}{0.5\textwidth}
        \centering
        \includegraphics[width=\linewidth]{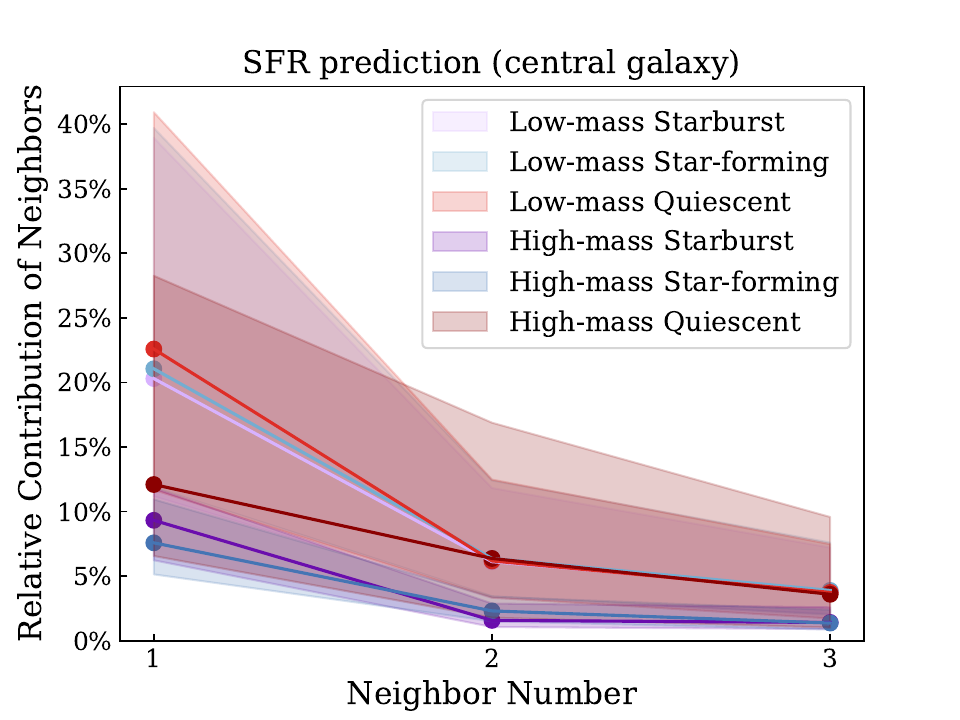}
    \end{minipage}
    \begin{minipage}{0.5\textwidth}
        \centering
        \includegraphics[width=\linewidth]{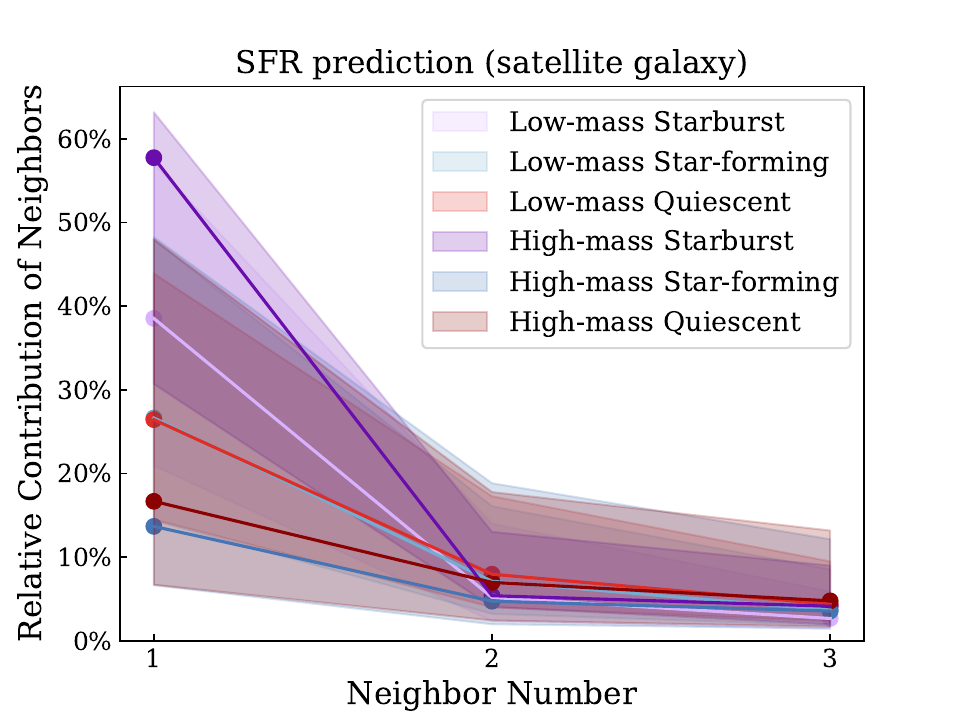}
    \end{minipage}
    \caption{
    The relative contributions of individual neighboring galaxies to the SFR prediction using the three-neighbors model, shown for central and satellite galaxies.
    These panels show the breakdown by the neighboring galaxy of the environmental contribution fractions presented in Fig.~\ref{tab:shap_frac_no}.
    Symbols indicate median values for each neighbor, and shaded regions represent the 16th to 84th percentile ranges.
    Among the three neighboring galaxies, the nearest neighbor contributes the most to the prediction.
    Results binned by distance on the horizontal axis are indicated in Fig.~\ref{Fig.D_bin_no}.
    }
    \label{Fig.SHAP_3Neighbor_bin}
\end{figure}

Fig.~\ref{Fig.SHAP_3Neighbor_bin} shows the relative contributions of individual neighboring galaxies to SFR predictions in our three-neighbors model. The results reveal a clear hierarchical pattern of environmental influence, with the nearest neighbor consistently providing the dominant contribution to SFR predictions across all galaxy types ($\sim$8-23\% for centrals and $\sim$14-58\% for satellites). This primary neighbor contribution is particularly pronounced for satellite starburst galaxies. The second-nearest neighbor contributes substantially less (<10\%), while the third neighbor offers minimal additional predictive information (typically below 5\%). The pattern holds across all galaxy mass ranges and star-formation states, indicating that while the overall magnitude of environmental dependence varies with galaxy types, the relative importance of neighbors follows a universal distance-based hierarchy. This is not a contradiction to the peak SFR prediction performance for the three-neighbors model. 
While the systematic reduction in information content with neighbor distance suggests a simple relationship where the environmental influence diminishes rapidly with separation, 
including additional neighbors helps average out stochasticity, and thus provides robustness in SFR predictions,
as later discussed in Section \ref{sec:discuss:environment}. 

\section{Discussion} \label{sec:discuss}
\subsection{Characterizing a proxy for the environment}\label{sec:discuss:environment}
Traditional approaches to quantifying galaxy environments typically employ spherical overdensities, fixed-scale density measurements, or cosmic web classifications (see Section~\ref{sec:intro}). 
Our approach differs fundamentally by parameterizing the environment through the $N$ nearest neighbors and their attributes, providing a flexible framework for isolating environmental contributions to galaxy properties. 
This neighbor-based approach naturally accommodates radial distance and subhalo properties of individual neighbors that are not captured by spherically-averaged metrics. 

As demonstrated in Fig.~\ref{Fig.violin}, prediction accuracy increases significantly when incorporating information from neighboring galaxies for both $M_*$ and SFR. While the exact optimal neighbor count shows modest variation within error bars, we identify peaks at one neighbor for $M_*$ and three neighbors for SFR. 

Though within the error bars, we see a tendency for the prediction accuracy to decrease when considering galaxies further away. 
This behavior suggests two possible explanations. 
The first possibility involves model training issues: increasing the number of nearby galaxies as inputs might have significantly expanded the input dimensionality of the MLP, complicating the learning process. If this is the case, enhancing model complexity and augmenting the training dataset would be necessary. 
The second possibility points toward inherent physical connections between galaxies. 
The model may have struggled when attempting to incorporate information from distant galaxies, whose physical influence on the target galaxy is likely negligible, resulting in reduced predictive performance. Although this study cannot entirely rule out either scenario, the consistent structure of the datasets, identical input feature types, and matching internal and hyperparameter settings across both the $M_*$ and SFR predictions lend weight to the second possibility. 

Thus, this optimal neighbor counts for $M_*$ and SFR reflects the distinct physical mechanisms and timescales governing these properties. 
For $M_*$ prediction, the dominance of the nearest neighbor reflects the cumulative nature of mass assembly.
$M_*$ integrates star formation history (SFH) over the assembly history lifetime, responding primarily to long-term gravitational interactions with the massive nearby perturber. 
The distance to the nearest neighbor ($D_1$) serves as a proxy for local environmental density: when $D_1$ is small, subsequent neighbors also tend to be nearby (for details, see \ref{sec:D_Q_app}), indicating that the galaxy resides in a dense environment with higher mass accretion rates. 
This suggests that the nearest neighbor encodes sufficient information about the local density field and its correlation with mass assembly history, such that including more distant neighbors does not improve $M_*$ predictions. 
This lack of improvement implies that higher-order environmental effects (from more distant neighbors) are either negligible or already encoded in the properties of the nearest neighbor through multiscaled correlated structure formation.

In contrast, SFR captures instantaneous gas conversion regulated by recent hydrodynamical interactions, gas accretion events, and feedback processes operating on multiple spatial scales. 
While $M_*$ is well constrained by the SHMR when $M_h$ is given, SFR exhibits larger fluctuations due to its sensitivity to short-timescale baryonic processes, making it more difficult to predict than $M_*$.
For SFR prediction, incorporating multiple neighbors effectively averages over the stochasticity of instantaneous gas dynamics, producing more robust predictions. 
The improved predictive power when incorporating three neighbors for SFR likely also reflects the complex nature of gas regulation: the nearest neighbor predominantly influences tidal torques and ram pressure stripping, while more distant neighbors affect gas supply through large-scale environments. 
The need for multiple neighbors in SFR prediction further suggests that gas regulation involves collective environmental effects that may not be captured by a single interaction.

\subsection{Isolating the influence from environment} \label{sec:discuss_no-assembly}
Disentangling pure environmental contributions from assembly history effects presents a significant methodological challenge. SHAP values quantify each input feature's contribution to predictions; however, they reflect correlation rather than causation. Features with high SHAP values do not necessarily directly impact galaxy mass or star formation but may instead correlate indirectly through other factors. Therefore, careful interpretation is essential to avoid misattributing correlation as causation. 

To address this, we construct models explicitly excluding $V_{\mathrm{max}}$ and $r_{50}$, which encode information related to halo and galaxy assembly history, from input features listed in Tables~\ref{tab:input_target} and \ref{tab:input_neighbors}, and analyze the resulting SHAP values. $V_{\mathrm{max}}$, representing the halo's gravitational potential depth, strongly correlates with $M_*$ and complicates SHAP interpretation. Additionally, $V_{\mathrm{max}}$ depends on halo formation timing and mass concentration, reflecting both environmental and assembly-history influences. Similarly, $r_{50}$ depends on gas accretion, feedback, and merger history, complicating the separation of environmental from internal factors. Thus, excluding $V_{\mathrm{max}}$ and $r_{50}$ is crucial for accurately evaluating environmental contributions. 
We retain $M_h$ as a parameter, as we are fundamentally interested in exploring how the secondary halo bias appears through the environment. 

As shown in Fig.~\ref{tab:shap_frac_no}, the contribution of environmental factors to predictions of $M_*$ and SFR varies significantly by galaxy type. Low-mass galaxies tend to experience greater environmental influence than high-mass galaxies, with satellite galaxies generally more susceptible than central galaxies.

One possible explanation is that high-mass galaxies and central galaxies, due to deeper gravitational potentials, are less susceptible to external environmental influences. Furthermore, high-mass galaxies frequently host AGNs, whose feedback can heat surrounding gas, thereby suppressing accretion. Consequently, in high-mass galaxies, gas availability likely relies more on internal structures (e.g., gas disks, molecular clouds) than external factors.

On the other hand, 
for low-mass central galaxies, the environmental contribution approaches $\sim$40\%, challenging the models where centrals evolve primarily through in-situ processes. This substantial environmental component for centrals aligns with evidence for large-scale tidal field influences on gas accretion \citep{hearinPhysicalOriginGalactic2016}.

Low-mass galaxies and satellite galaxies, possessing shallower gravitational potentials, are more influenced by external environmental effects such as large-scale structures, tidal interactions, and ram pressure stripping. Such influences may strip gas and quench star formation or compress gas to boost star formation temporarily. Indeed, in satellite starburst galaxies, environmental factors contribute more significantly to SFR predictions in high-mass galaxies, suggesting deeper gravitational potentials enhance gas accretion and star formation. These results indicate that external environments do not necessarily exert a consistent influence on star formation. Given that $M_*$ reflects cumulative SFH, long-term trends in $M_*$ predictions should correspond closely with those observed in SFR.

From the SHAP analysis, the environmental factors that most significantly contribute to predicting $M_*$ and SFR are the distance $D$ to neighboring galaxies, their relative velocity $v_D$, and the halo mass $M_h$ of surrounding galaxies, as shown in Fig.~\ref{Fig.SHAP_feature_no}.
The importance of $D$ across all populations emphasizes the inverse-square scaling of gravitational and hydrodynamical interactions. 
In particular, satellite starburst galaxies exhibit particularly high SHAP values for $D$, suggesting that interactions with nearby galaxies play a crucial role in controlling physical processes within galaxies. Tidal forces may compress gas, enhancing star formation, or interactions might strip gas, suppressing star formation. Additionally, the significant contribution of $v_D$ highlights the importance of dynamical interactions, with large $v_D$ potentially increasing gas removal through ram pressure stripping in dense environments such as galaxy clusters. 
Surrounding $M_h$ likely informs the model regarding central versus satellite galaxy classification, enhancing potential-related information and thus improving the accuracy of $M_*$ predictions.

\subsection{Implications for galaxy formation models}
Our results provide guidance for improving galaxy formation models, particularly for satellites, low-mass systems, and starburst populations. Current empirical models like \textsc{UniverseMachine} \citep{behrooziUniverseMachineCorrelationGalaxy2019} and 
EMERGE \citep{mosterEmergeEmpiricalModel2018}
primarily employ assembly history as a secondary parameter modulating galaxy properties. With our findings, we suggest specific implementable improvements through neighbor-based environmental metrics.

First, empirical models could adopt mass-dependent and type-dependent environmental efficiency functions rather than universal prescriptions. For centrals, environmental influence should scale inversely with mass, approaching zero for $\log(M_*/M_{\odot}) > 10$ while remaining substantial ($\sim$40\%) for $\log(M_*/M_{\odot}) < 10$. For satellites, environmental influence should maintain $\sim$40-50\% contribution across mass ranges but increase specifically for massive starbursts.

Second, the differential environmental scale dependence for $M_*$ versus SFR suggests that multi-scale environmental treatments would significantly improve model accuracy. $M_*$ assignment could employ nearest-neighbor metrics, while SFR assignment should incorporate broader environmental information extending to at least the third nearest neighbor. Having more than one neighbor in the prediction provides unique additional information.

This approach appears particularly crucial for modeling certain subpopulations that have been challenging with current models. Emission-line galaxies (ELGs) often represent merger-triggered starbursts with disturbed morphologies, whose star formation is driven by tidal interactions rather than secular evolution or assembly history \citep{yuanUnravelingEmissionLine2025}. Modeling observed enhanced quenching fractions of dwarf satellites relative to the isolated counterparts could also benefit from this additional environment-dependent degree of freedom not seen for massive systems \citep{WangY_2024}. A neighbor-based parameterization provides a computationally efficient and natural way to model this merger-driven star formation without requiring full hydrodynamical treatment.

For semi-analytic models, our findings suggest that a form of "communication" between distinct merger trees would significantly improve accuracy, especially for star formation predictions. Current implementations typically evolve individual halos in isolation except during explicit mergers, neglecting pre-merger interactions that our analysis shows contribute substantially to galaxy properties. Implementing interaction physics based on neighbor distances and velocities prior to mergers would better capture ELG-like populations with interaction-triggered star formation. While these effects may be difficult to implement in the set of governing equations, new approaches that mimic semi-analytical predictions, e.g., \citet{Jesperson_2022} may be extended to include such effects.

Importantly, these suggested modifications are observationally accessible through pair counts and local density measurements, providing empirical proxies for assembly bias effects that can be directly constrained through observed merger pairs \citep{Behroozi2015}. Additionally, as demonstrated by \citet{yuanUnravelingEmissionLine2025}, we could understand more about environment-driven processes through cross-correlation analyses of different galaxy subpopulations.

The environmental dependencies quantified in our work, particularly the different scales for $M_*$ versus SFR prediction, offer a powerful framework for improving theoretical models.
By capturing the neighbor-dependent enhancement of star formation seen observationally in interacting systems, our approach addresses a key limitation in current empirical models.
For simulations with limited resolution or without merger trees, our quantification of the necessary scales for environmental influence provides guidance on the modeling required to accurately reproduce all galaxy populations.

\subsection{Methodological limitations and future directions}
While our neural network approach effectively quantifies environmental contributions, several limitations constrain interpretation. Our $z=0$ analysis provides a single-time snapshot view rather than tracking the evolving galaxy-environment relationship. Environmental effects operate cumulatively across cosmic time, with present-day properties potentially reflecting historical conditions not captured in current neighbor configurations.

The prediction accuracy varies significantly across the galaxy population, with notable poor prediction for quiescent galaxies, particularly at low masses. This likely reflects both the stochastic nature of implemented star formation in quenched galaxies and the accumulated influence of historical environmental conditions not captured in present-day neighbor properties.
In fact, some previous studies \citep[e.g.,][]{Conroy_2006, Moster_2010, Moster_2013, Moster_2021} pointed out that peak halo mass over its history ($M_{\mathrm{peak}}$), or $V_{\mathrm{max}}$ when $M_{\mathrm{peak}}$ is achieved are strong predictors of the $M_*$.

At present, our model performance is improved by taking account of $V_{\mathrm{max}}$ and $r_{50}$, which indirectly capture some aspects of the assembly history. 
Nevertheless, incorporating a more detailed assembly history, such as through a merger tree, would likely enhance the accuracy of galaxy property predictions.
Moreover, in \cite{Montero_2021}, they tracked the progenitors of present galaxies and found that the signal of assembly bias is significantly enhanced at higher redshifts (e.g., $z=1$) compared to $z=0$. 
Therefore, when predicting galaxy properties at high redshift, it may be more important to fully account for the assembly history.
Extending our analysis to multiple redshifts would characterize the evolving importance of environmental regulation throughout cosmic history, potentially revealing distinct environmental regimes that could be incorporated into next-generation empirical models.

One of the limitations of our study is the simulation volume that we use.
Althogh TNG300-1 is one of the largest high-resolution hydrodynamical simulation, it relatively small cosmological volume. 
This may constrains our ability to capture the full diversity of extreme environmental conditions, especially the most overdense and underdense regions, and limits the number of very massive galaxies ($M_* > 10^{11} M_{\odot}$) in our sample.
Future work using larger-volume, higher-resolution hydrodynamical simulations, analyzed through our approach, will allow for more comprehensive characterization of environmental diversity and enable analyses across a wider range of galaxy masses. 

Furthermore, by applying our methodology to other simulation suites such as the CAMELS \citep{Villaescusa-Navarro2021} and AGORA \citep{Kim2016} suites, it becomes possible to assess the robustness of our findings against variations in physical prescriptions, either due to differences in subgrid physics or differences in how similar subgrid physics are implemented across simulations.
The CAMELS offers a testbed with its systematic variation of cosmological and astrophysical parameters across a lot of simulations, allowing us to assess how different feedback implementations affect the relative importance of environmental factors. 
While the AGORA provides fewer simulations, its focus on high-resolution galaxy formation with diverse numerical methods offers complementary insights into how numerical implementation choices might influence our conclusions about environmental effects. 

\section{Conclusions} \label{sec:conclusions}
In this study,  we have developed an interpretable machine learning framework to quantify how galaxy properties depend on their surrounding environment, using the IllustrisTNG cosmological hydrodynamical simulation. By training neural networks to predict stellar mass ($M_*$) and star formation rates (SFR) from dark matter (DM) halo properties, we isolate and measure environmental influences across different galaxy populations. Our key findings are:

\begin{enumerate}
    \item Both $M_*$ and SFR predictions improve statistically significantly when incorporating information from neighboring DM halos, confirming that environmental effects are critical for accurately modeling galaxy populations.

    \item Central and satellite galaxies experience fundamentally different environmental effects. Satellites consistently show 40-50\% environmental contribution regardless of the mass bin, whereas centrals exhibit mass-dependent environmental sensitivity.
    
    \item Environmental dependence exhibits a strong inverse correlation with mass (a proxy for potential well depth) for centrals. 
    Low-mass galaxies are more strongly influenced by environmental factors.
    
    \item The number of neighbors needed for optimal prediction differs between $M_*$ and SFR: $M_*$ requires only the nearest neighbor (reflecting integrated gravitational effects), while SFR benefits from information up to the third-nearest neighbor (capturing complex gas regulation mechanisms).
    
    \item Neighbor contribution follows a clear hierarchy on average, with the first neighbor providing the dominant contribution, the second neighbor offering diminishing returns, and the third neighbor providing minimal additional information.
    
    \item From the neighbor attributes, their distance to the target halo consistently emerges as the most significant environmental predictor across all galaxy populations, followed by relative velocity for $M_*$ predictions and neighbor halo mass for SFR predictions. 
    
    \item Different galaxy subpopulations (based on star formation activity and mass) exhibit distinct environmental dependencies, with satellite starbursts showing anomalously high environmental sensitivity despite their massive system.
\end{enumerate}

Our results have significant implications for galaxy formation modeling. Current empirical and semi-analytic models must incorporate explicit environmental dependency that accounts for satellite/central status, galaxy mass, and star formation classification. Simply using $M_h$ and assembly history is insufficient to fully capture the environmental effects we have quantified. Particularly for emission-line galaxies, dwarf satellites, and quiescent populations, neighbor-based metrics provide essential information that assembly history alone cannot encode.

A key physical aspect to note is the timescale of environmental influence. Environmental effects can manifest on short timescales, such as gas supply and removal resulting from interactions, or over longer timescales, such as continuous quenching following a galaxy's transition to a satellite, or gas supply shortages in low-mass halos that shape the SFH. Since our study does not account for temporal variations, incorporating past environmental histories through the use of merger trees represents an important direction for future research.

Our quantitative framework for environmental dependence enables more sophisticated galaxy-halo connection models that can capture the diversity of galaxy populations observed across different environments. By incorporating the neighbor-dependent modulation of star formation seen in interacting systems and systematically accounting for mass-dependent environmental sensitivity, future empirical models can better reproduce observed 1-point and 2-point statistics of galaxy populations while remaining computationally efficient for large-volume cosmological applications.

\section*{Acknowledgements}
This work has been supported by the Japan Society for the Promotion of Science (JSPS) Grants-in-Aid for Scientific Research, 24H00247, 25K01044, 25H01551, 23H00108, 22K21349 and 21K03625.
This work has also been supported in part by the Collaboration Funding of the Institute of Statistical Mathematics ``Machine-Learning-Based Cosmogony: From Structure Formation to Galaxy Evolution''. 

Numerical computations were in part carried out on GPU cluster at the Center for Computational Astrophysics, National Astronomical Observatory of Japan.

We utilized the following Python libraries: NumPy \citep{Numpy}, Matplotlib \citep{matplotlib}, SciencePlots \citep{SciencePlots}, Astropy \citep{astropy:2022}, pandas \citep{reback2020pandas}, scikit-learn \citep{sklearn}, SciPy \citep{SciPy}, PyTorch \citep{pytorch}, and SHAP \citep{SHAP_NIPS2017_7062}, among others.

\section*{Data Availability}
The IllustrisTNG simulation data is accessible in \href{https://www.tng-project.org/data}{https://www.tng-project.org/data}. 
The models and analysis code in this paper will be shared upon reasonable request to the authors.
 


\bibliographystyle{mnras}
\bibliography{reference} 

\begin{thebibliography}{}
\makeatletter
\relax
\def\mn@urlcharsother{\let\do\@makeother \do\$\do\&\do\#\do\^\do\_\do\%\do\~}
\def\mn@doi{\begingroup\mn@urlcharsother \@ifnextchar [ {\mn@doi@} {\mn@doi@[]}}
\def\mn@doi@[#1]#2{\def\@tempa{#1}\ifx\@tempa\@empty \href {http://dx.doi.org/#2} {doi:#2}\else \href {http://dx.doi.org/#2} {#1}\fi \endgroup}
\def\mn@eprint#1#2{\mn@eprint@#1:#2::\@nil}
\def\mn@eprint@arXiv#1{\href {http://arxiv.org/abs/#1} {{\tt arXiv:#1}}}
\def\mn@eprint@dblp#1{\href {http://dblp.uni-trier.de/rec/bibtex/#1.xml} {dblp:#1}}
\def\mn@eprint@#1:#2:#3:#4\@nil{\def\@tempa {#1}\def\@tempb {#2}\def\@tempc {#3}\ifx \@tempc \@empty \let \@tempc \@tempb \let \@tempb \@tempa \fi \ifx \@tempb \@empty \def\@tempb {arXiv}\fi \@ifundefined {mn@eprint@\@tempb}{\@tempb:\@tempc}{\expandafter \expandafter \csname mn@eprint@\@tempb\endcsname \expandafter{\@tempc}}}

\bibitem[\protect\citeauthoryear{{Agarwal}, {Dav{\'e}}  \& {Bassett}}{{Agarwal} et~al.}{2018}]{Agarwal_2018}
{Agarwal} S.,  {Dav{\'e}} R.,   {Bassett} B.~A.,  2018, \mn@doi [\mnras] {10.1093/mnras/sty1169}, \href {https://ui.adsabs.harvard.edu/abs/2018MNRAS.478.3410A} {478, 3410}

\bibitem[\protect\citeauthoryear{{Aguilar-Arg{\"u}ello} et~al.,}{{Aguilar-Arg{\"u}ello} et~al.}{2025}]{2025MNRAS.537..876A}
{Aguilar-Arg{\"u}ello} G.,  et~al., 2025, \mn@doi [\mnras] {10.1093/mnras/staf085}, \href {https://ui.adsabs.harvard.edu/abs/2025MNRAS.537..876A} {537, 876}

\bibitem[\protect\citeauthoryear{Alfonzo et~al.,}{Alfonzo et~al.}{2024}]{alfonzoKatachiXingDecoding2024}
Alfonzo J.~P.,  et~al., 2024, \mn@doi [\apj] {10.3847/1538-4357/ad3b95}, 967, 152

\bibitem[\protect\citeauthoryear{{Astropy Collaboration} et~al.,}{{Astropy Collaboration} et~al.}{2022}]{astropy:2022}
{Astropy Collaboration} et~al., 2022, \mn@doi [\apj] {10.3847/1538-4357/ac7c74}, \href {https://ui.adsabs.harvard.edu/abs/2022ApJ...935..167A} {935, 167}

\bibitem[\protect\citeauthoryear{Ayromlou, Kauffmann, Anand  \& White}{Ayromlou et~al.}{2022}]{ayromlouPhysicalOriginGalactic2022}
Ayromlou M.,  Kauffmann G.,  Anand A.,   White S. D.~M.,  2022, \mn@doi [\mnras] {10.1093/mnras/stac3637}, 519, 1913

\bibitem[\protect\citeauthoryear{{Ayromlou}, {Kauffmann}, {Anand}  \& {White}}{{Ayromlou} et~al.}{2023}]{Ayromlou_2023}
{Ayromlou} M.,  {Kauffmann} G.,  {Anand} A.,   {White} S. D.~M.,  2023, \mn@doi [\mnras] {10.1093/mnras/stac3637}, \href {https://ui.adsabs.harvard.edu/abs/2023MNRAS.519.1913A} {519, 1913}

\bibitem[\protect\citeauthoryear{{Banerjee} \& {Abel}}{{Banerjee} \& {Abel}}{2021}]{Banerjee_2021}
{Banerjee} A.,  {Abel} T.,  2021, \mn@doi [\mnras] {10.1093/mnras/staa3604}, \href {https://ui.adsabs.harvard.edu/abs/2021MNRAS.500.5479B} {500, 5479}

\bibitem[\protect\citeauthoryear{Behroozi, Conroy  \& Wechsler}{Behroozi et~al.}{2010}]{Behroozi_2010}
Behroozi P.~S.,  Conroy C.,   Wechsler R.~H.,  2010, \mn@doi [\apj] {10.1088/0004-637X/717/1/379}, 717, 379

\bibitem[\protect\citeauthoryear{{Behroozi}, {Wechsler}  \& {Conroy}}{{Behroozi} et~al.}{2013}]{2013ApJ...770...57B}
{Behroozi} P.~S.,  {Wechsler} R.~H.,   {Conroy} C.,  2013, \mn@doi [\apj] {10.1088/0004-637X/770/1/57}, \href {https://ui.adsabs.harvard.edu/abs/2013ApJ...770...57B} {770, 57}

\bibitem[\protect\citeauthoryear{{Behroozi} et~al.,}{{Behroozi} et~al.}{2015}]{Behroozi2015}
{Behroozi} P.~S.,  et~al., 2015, \mn@doi [\mnras] {10.1093/mnras/stv728}, \href {https://ui.adsabs.harvard.edu/abs/2015MNRAS.450.1546B} {450, 1546}

\bibitem[\protect\citeauthoryear{Behroozi, Wechsler, Hearin  \& Conroy}{Behroozi et~al.}{2019}]{behrooziUniverseMachineCorrelationGalaxy2019}
Behroozi P.,  Wechsler R.~H.,  Hearin A.~P.,   Conroy C.,  2019, \mn@doi [\mnras] {10.1093/mnras/stz1182}, 488, 3143

\bibitem[\protect\citeauthoryear{{Blanton}, {Eisenstein}, {Hogg}, {Schlegel}  \& {Brinkmann}}{{Blanton} et~al.}{2005}]{Blanton_2005}
{Blanton} M.~R.,  {Eisenstein} D.,  {Hogg} D.~W.,  {Schlegel} D.~J.,   {Brinkmann} J.,  2005, \mn@doi [\apj] {10.1086/422897}, \href {https://ui.adsabs.harvard.edu/abs/2005ApJ...629..143B} {629, 143}

\bibitem[\protect\citeauthoryear{{Blanton}, {Eisenstein}, {Hogg}  \& {Zehavi}}{{Blanton} et~al.}{2006}]{Blanton_2006}
{Blanton} M.~R.,  {Eisenstein} D.,  {Hogg} D.~W.,   {Zehavi} I.,  2006, \mn@doi [\apj] {10.1086/500918}, \href {https://ui.adsabs.harvard.edu/abs/2006ApJ...645..977B} {645, 977}

\bibitem[\protect\citeauthoryear{Camargo \& Casas-Miranda}{Camargo \& Casas-Miranda}{2025}]{camargo2025}
Camargo Y.~D.,  Casas-Miranda R.~A.,  2025, \mn@doi [\mnras] {10.1093/mnras/staf272}, 538, 312

\bibitem[\protect\citeauthoryear{{Chuang}, {Jespersen}, {Lin}, {Ho}  \& {Genel}}{{Chuang} et~al.}{2024}]{Chuang_2024}
{Chuang} C.-Y.,  {Jespersen} C.~K.,  {Lin} Y.-T.,  {Ho} S.,   {Genel} S.,  2024, \mn@doi [\apj] {10.3847/1538-4357/ad2b6c}, \href {https://ui.adsabs.harvard.edu/abs/2024ApJ...965..101C} {965, 101}

\bibitem[\protect\citeauthoryear{Conroy, Wechsler  \& Kravtsov}{Conroy et~al.}{2006}]{Conroy_2006}
Conroy C.,  Wechsler R.~H.,   Kravtsov A.~V.,  2006, \mn@doi [\apj] {10.1086/503602}, 647, 201

\bibitem[\protect\citeauthoryear{{Croton}, {Gao}  \& {White}}{{Croton} et~al.}{2007}]{Croton_2007}
{Croton} D.~J.,  {Gao} L.,   {White} S. D.~M.,  2007, \mn@doi [\mnras] {10.1111/j.1365-2966.2006.11230.x}, \href {https://ui.adsabs.harvard.edu/abs/2007MNRAS.374.1303C} {374, 1303}

\bibitem[\protect\citeauthoryear{{Dalal}, {White}, {Bond}  \& {Shirokov}}{{Dalal} et~al.}{2008}]{Dalal_2008}
{Dalal} N.,  {White} M.,  {Bond} J.~R.,   {Shirokov} A.,  2008, \mn@doi [\apj] {10.1086/591512}, \href {https://ui.adsabs.harvard.edu/abs/2008ApJ...687...12D} {687, 12}

\bibitem[\protect\citeauthoryear{Donnari et~al.,}{Donnari et~al.}{2019}]{donnariStarFormationActivity2019a}
Donnari M.,  et~al., 2019, \mn@doi [\mnras] {10.1093/mnras/stz712}, 485, 4817

\bibitem[\protect\citeauthoryear{Dressler}{Dressler}{1980}]{dresslerGalaxyMorphologyRich1980}
Dressler A.,  1980, \mn@doi [\apj] {10.1086/157753}, 236, 351

\bibitem[\protect\citeauthoryear{{Gal{\'a}rraga-Espinosa}, {Langer}  \& {Aghanim}}{{Gal{\'a}rraga-Espinosa} et~al.}{2022}]{Galarraga_2022}
{Gal{\'a}rraga-Espinosa} D.,  {Langer} M.,   {Aghanim} N.,  2022, \mn@doi [\aap] {10.1051/0004-6361/202141974}, \href {https://ui.adsabs.harvard.edu/abs/2022A&A...661A.115G} {661, A115}

\bibitem[\protect\citeauthoryear{{Gao}, {Springel}  \& {White}}{{Gao} et~al.}{2005}]{Gao2005}
{Gao} L.,  {Springel} V.,   {White} S. D.~M.,  2005, \mn@doi [\mnras] {10.1111/j.1745-3933.2005.00084.x}, \href {https://ui.adsabs.harvard.edu/abs/2005MNRAS.363L..66G} {363, L66}

\bibitem[\protect\citeauthoryear{Garrett}{Garrett}{2021}]{SciencePlots}
Garrett J.~D.,  2021, \mn@doi{10.5281/zenodo.4106649}

\bibitem[\protect\citeauthoryear{{Girelli, G.}, {Pozzetti, L.}, {Bolzonella, M.}, {Giocoli, C.}, {Marulli, F.}  \& {Baldi, M.}}{{Girelli, G.} et~al.}{2020}]{SHMR}
{Girelli, G.} {Pozzetti, L.} {Bolzonella, M.} {Giocoli, C.} {Marulli, F.}  {Baldi, M.} 2020, \mn@doi [A&A] {10.1051/0004-6361/201936329}, 634, A135

\bibitem[\protect\citeauthoryear{{Gunn} \& {Gott}}{{Gunn} \& {Gott}}{1972}]{ram_1}
{Gunn} J.~E.,  {Gott} III J.~R.,  1972, \mn@doi [\apj] {10.1086/151605}, \href {https://ui.adsabs.harvard.edu/abs/1972ApJ...176....1G} {176, 1}

\bibitem[\protect\citeauthoryear{Guo et~al.,}{Guo et~al.}{2011}]{1halo_sim}
Guo Q.,  et~al., 2011, \mn@doi [\mnras] {10.1111/j.1365-2966.2010.18114.x}, 413, 101

\bibitem[\protect\citeauthoryear{{Hahn}, {Porciani}, {Dekel}  \& {Carollo}}{{Hahn} et~al.}{2009}]{Hahn_2009}
{Hahn} O.,  {Porciani} C.,  {Dekel} A.,   {Carollo} C.~M.,  2009, \mn@doi [\mnras] {10.1111/j.1365-2966.2009.15271.x}, \href {https://ui.adsabs.harvard.edu/abs/2009MNRAS.398.1742H} {398, 1742}

\bibitem[\protect\citeauthoryear{Harris et~al.,}{Harris et~al.}{2020}]{Numpy}
Harris C.~R.,  et~al., 2020, \mn@doi [Nature] {10.1038/s41586-020-2649-2}, 585, 357

\bibitem[\protect\citeauthoryear{{Hasan} et~al.,}{{Hasan} et~al.}{2024}]{Hasan_2024}
{Hasan} F.,  et~al., 2024, \mn@doi [\apj] {10.3847/1538-4357/ad4ee2}, \href {https://ui.adsabs.harvard.edu/abs/2024ApJ...970..177H} {970, 177}

\bibitem[\protect\citeauthoryear{{He}, {Luo}  \& {Chen}}{{He} et~al.}{2022}]{2022MNRAS.512.1710H}
{He} X.-J.,  {Luo} A.~L.,   {Chen} Y.-Q.,  2022, \mn@doi [\mnras] {10.1093/mnras/stac484}, \href {https://ui.adsabs.harvard.edu/abs/2022MNRAS.512.1710H} {512, 1710}

\bibitem[\protect\citeauthoryear{{Hearin}, {Watson}  \& {van den Bosch}}{{Hearin} et~al.}{2015}]{Hearin_2015}
{Hearin} A.~P.,  {Watson} D.~F.,   {van den Bosch} F.~C.,  2015, \mn@doi [\mnras] {10.1093/mnras/stv1358}, \href {https://ui.adsabs.harvard.edu/abs/2015MNRAS.452.1958H} {452, 1958}

\bibitem[\protect\citeauthoryear{Hearin, Behroozi  \& Van Den~Bosch}{Hearin et~al.}{2016}]{hearinPhysicalOriginGalactic2016}
Hearin A.~P.,  Behroozi P.~S.,   Van Den~Bosch F.~C.,  2016, \mn@doi [\mnras] {10.1093/mnras/stw1462}, 461, 2135

\bibitem[\protect\citeauthoryear{{Hogg} et~al.,}{{Hogg} et~al.}{2003}]{Hogg_2003}
{Hogg} D.~W.,  et~al., 2003, \mn@doi [\apjl] {10.1086/374238}, \href {https://ui.adsabs.harvard.edu/abs/2003ApJ...585L...5H} {585, L5}

\bibitem[\protect\citeauthoryear{Hopkins, Hernquist, Cox, Matteo, Robertson  \& Springel}{Hopkins et~al.}{2006}]{hopkinsUnifiedMergerdrivenModel2006}
Hopkins P.~F.,  Hernquist L.,  Cox T.~J.,  Matteo T.~D.,  Robertson B.,   Springel V.,  2006, \mn@doi [ApJS] {10.1086/499298}, 163, 1

\bibitem[\protect\citeauthoryear{Hunter}{Hunter}{2007}]{matplotlib}
Hunter J.~D.,  2007, \mn@doi [Computing in Science & Engineering] {10.1109/MCSE.2007.55}, 9, 90

\bibitem[\protect\citeauthoryear{Iwasaki, Cooray  \& Takeuchi}{Iwasaki et~al.}{2023}]{iwasakiExtractingInformativeLatent2023}
Iwasaki D.,  Cooray S.,   Takeuchi T.~T.,  2023, {Extracting an {{Informative Latent Representation}} of {{High-Dimensional Galaxy Spectra}}} (\mn@eprint {arXiv} {2311.17414})

\bibitem[\protect\citeauthoryear{{Jespersen}, {Cranmer}, {Melchior}, {Ho}, {Somerville}  \& {Gabrielpillai}}{{Jespersen} et~al.}{2022}]{Jesperson_2022}
{Jespersen} C.~K.,  {Cranmer} M.,  {Melchior} P.,  {Ho} S.,  {Somerville} R.~S.,   {Gabrielpillai} A.,  2022, \mn@doi [\apj] {10.3847/1538-4357/ac9b18}, \href {https://ui.adsabs.harvard.edu/abs/2022ApJ...941....7J} {941, 7}

\bibitem[\protect\citeauthoryear{{Kamdar}, {Turk}  \& {Brunner}}{{Kamdar} et~al.}{2016}]{Kamdar_2016}
{Kamdar} H.~M.,  {Turk} M.~J.,   {Brunner} R.~J.,  2016, \mn@doi [\mnras] {10.1093/mnras/stv2981}, \href {https://ui.adsabs.harvard.edu/abs/2016MNRAS.457.1162K} {457, 1162}

\bibitem[\protect\citeauthoryear{Kauffmann, Li, Zhang  \& Weinmann}{Kauffmann et~al.}{2013}]{kauffmannReexaminationGalacticConformity2013}
Kauffmann G.,  Li C.,  Zhang W.,   Weinmann S.,  2013, \mn@doi [\mnras] {10.1093/mnras/stt007}, 430, 1447

\bibitem[\protect\citeauthoryear{{Kim} et~al.,}{{Kim} et~al.}{2016}]{Kim2016}
{Kim} J.-h.,  et~al., 2016, \mn@doi [\apj] {10.3847/1538-4357/833/2/202}, \href {https://ui.adsabs.harvard.edu/abs/2016ApJ...833..202K} {833, 202}

\bibitem[\protect\citeauthoryear{{Kono}, {Takeuchi}, {Cooray}, {Nishizawa}  \& {Murakami}}{{Kono} et~al.}{2020}]{Kono_2020}
{Kono} K.~T.,  {Takeuchi} T.~T.,  {Cooray} S.,  {Nishizawa} A.~J.,   {Murakami} K.,  2020, \mn@doi [arXiv e-prints] {10.48550/arXiv.2006.02905}, \href {https://ui.adsabs.harvard.edu/abs/2020arXiv200602905K} {}

\bibitem[\protect\citeauthoryear{{Lacerna}, {Contreras}, {Gonz{\'a}lez}, {Padilla}  \& {Gonzalez-Perez}}{{Lacerna} et~al.}{2018}]{lacerna_2018}
{Lacerna} I.,  {Contreras} S.,  {Gonz{\'a}lez} R.~E.,  {Padilla} N.,   {Gonzalez-Perez} V.,  2018, \mn@doi [\mnras] {10.1093/mnras/stx3253}, \href {https://ui.adsabs.harvard.edu/abs/2018MNRAS.475.1177L} {475, 1177}

\bibitem[\protect\citeauthoryear{Lacerna et~al.,}{Lacerna et~al.}{2022}]{lacernaEnvironmentalInfluenceGroups2022}
Lacerna I.,  et~al., 2022, \mn@doi [\mnras] {10.1093/mnras/stac1020}, 513, 2271

\bibitem[\protect\citeauthoryear{{Larson}, {Tinsley}  \& {Caldwell}}{{Larson} et~al.}{1980}]{strangulation1}
{Larson} R.~B.,  {Tinsley} B.~M.,   {Caldwell} C.~N.,  1980, \mn@doi [\apj] {10.1086/157917}, \href {https://ui.adsabs.harvard.edu/abs/1980ApJ...237..692L} {237, 692}

\bibitem[\protect\citeauthoryear{{Lim}, {Mo}, {Wang}  \& {Yang}}{{Lim} et~al.}{2016}]{Lim2016}
{Lim} S.~H.,  {Mo} H.~J.,  {Wang} H.,   {Yang} X.,  2016, \mn@doi [\mnras] {10.1093/mnras/stv2282}, \href {https://ui.adsabs.harvard.edu/abs/2016MNRAS.455..499L} {455, 499}

\bibitem[\protect\citeauthoryear{Lin et~al.,}{Lin et~al.}{2010}]{Lin_2010}
Lin L.,  et~al., 2010, \mn@doi [\apj] {10.1088/0004-637X/718/2/1158}, 718, 1158

\bibitem[\protect\citeauthoryear{Loshchilov \& Hutter}{Loshchilov \& Hutter}{2019}]{loshchilov2019decoupledweightdecayregularization}
Loshchilov I.,  Hutter F.,  2019, Decoupled Weight Decay Regularization (\mn@eprint {arXiv} {1711.05101})

\bibitem[\protect\citeauthoryear{Lundberg \& Lee}{Lundberg \& Lee}{2017}]{SHAP_NIPS2017_7062}
Lundberg S.,  Lee S.-I.,  2017, A Unified Approach to Interpreting Model Predictions (\mn@eprint {arXiv} {1705.07874})

\bibitem[\protect\citeauthoryear{{Mao}, {Zentner}  \& {Wechsler}}{{Mao} et~al.}{2018}]{Mao_2018}
{Mao} Y.-Y.,  {Zentner} A.~R.,   {Wechsler} R.~H.,  2018, \mn@doi [\mnras] {10.1093/mnras/stx3111}, \href {https://ui.adsabs.harvard.edu/abs/2018MNRAS.474.5143M} {474, 5143}

\bibitem[\protect\citeauthoryear{{Marinacci} et~al.,}{{Marinacci} et~al.}{2018}]{TNG3}
{Marinacci} F.,  et~al., 2018, \mn@doi [\mnras] {10.1093/mnras/sty2206}, \href {https://ui.adsabs.harvard.edu/abs/2018MNRAS.480.5113M} {480, 5113}

\bibitem[\protect\citeauthoryear{Molnar}{Molnar}{2022}]{molnarInterpretableMachineLearning}
Molnar C.,  2022, Interpretable {{Machine Learning}} ({{Second Edition}}).
Leanpub, \url {https://christophm.github.io/interpretable-ml-book/}

\bibitem[\protect\citeauthoryear{{Montero-Dorta} et~al.,}{{Montero-Dorta} et~al.}{2017}]{Montero2017}
{Montero-Dorta} A.~D.,  et~al., 2017, \mn@doi [\apjl] {10.3847/2041-8213/aa8cc5}, \href {https://ui.adsabs.harvard.edu/abs/2017ApJ...848L...2M} {848, L2}

\bibitem[\protect\citeauthoryear{{Montero-Dorta}, {Chaves-Montero}, {Artale}  \& {Favole}}{{Montero-Dorta} et~al.}{2021}]{Montero_2021}
{Montero-Dorta} A.~D.,  {Chaves-Montero} J.,  {Artale} M.~C.,   {Favole} G.,  2021, \mn@doi [\mnras] {10.1093/mnras/stab2556}, \href {https://ui.adsabs.harvard.edu/abs/2021MNRAS.508..940M} {508, 940}

\bibitem[\protect\citeauthoryear{Moore, Katz, Lake, Dressler  \& Oemler}{Moore et~al.}{1996}]{mooreGalaxyHarassmentEvolution1996}
Moore B.,  Katz N.,  Lake G.,  Dressler A.,   Oemler A.,  1996, \mn@doi [Nature] {10.1038/379613a0}, 379, 613

\bibitem[\protect\citeauthoryear{Moster, Somerville, Maulbetsch, van~den Bosch, Macciò, Naab  \& Oser}{Moster et~al.}{2010}]{Moster_2010}
Moster B.~P.,  Somerville R.~S.,  Maulbetsch C.,  van~den Bosch F.~C.,  Macciò A.~V.,  Naab T.,   Oser L.,  2010, \mn@doi [\apj] {10.1088/0004-637X/710/2/903}, 710, 903

\bibitem[\protect\citeauthoryear{{Moster}, {Naab}  \& {White}}{{Moster} et~al.}{2013}]{Moster_2013}
{Moster} B.~P.,  {Naab} T.,   {White} S. D.~M.,  2013, \mn@doi [\mnras] {10.1093/mnras/sts261}, \href {https://ui.adsabs.harvard.edu/abs/2013MNRAS.428.3121M} {428, 3121}

\bibitem[\protect\citeauthoryear{Moster, Naab  \& White}{Moster et~al.}{2018}]{mosterEmergeEmpiricalModel2018}
Moster B.~P.,  Naab T.,   White S. D.~M.,  2018, \mn@doi [\mnras] {10.1093/mnras/sty655}, 477, 1822

\bibitem[\protect\citeauthoryear{{Moster}, {Naab}, {Lindstr{\"o}m}  \& {O'Leary}}{{Moster} et~al.}{2021}]{Moster_2021}
{Moster} B.~P.,  {Naab} T.,  {Lindstr{\"o}m} M.,   {O'Leary} J.~A.,  2021, \mn@doi [\mnras] {10.1093/mnras/stab1449}, \href {https://ui.adsabs.harvard.edu/abs/2021MNRAS.507.2115M} {507, 2115}

\bibitem[\protect\citeauthoryear{{Naiman} et~al.,}{{Naiman} et~al.}{2018}]{TNG4}
{Naiman} J.~P.,  et~al., 2018, \mn@doi [\mnras] {10.1093/mnras/sty618}, \href {https://ui.adsabs.harvard.edu/abs/2018MNRAS.477.1206N} {477, 1206}

\bibitem[\protect\citeauthoryear{Nelson et~al.,}{Nelson et~al.}{2017}]{TNG2}
Nelson D.,  et~al., 2017, \mn@doi [\mnras] {10.1093/mnras/stx3040}, 475, 624

\bibitem[\protect\citeauthoryear{{Olsen} \& {Gawiser}}{{Olsen} \& {Gawiser}}{2023}]{Olsen_2023}
{Olsen} C.,  {Gawiser} E.,  2023, \mn@doi [\apj] {10.3847/1538-4357/acaa39}, \href {https://ui.adsabs.harvard.edu/abs/2023ApJ...943...30O} {943, 30}

\bibitem[\protect\citeauthoryear{{Olsen} et~al.,}{{Olsen} et~al.}{2021}]{Olsen_2021}
{Olsen} C.,  et~al., 2021, \mn@doi [\apj] {10.3847/1538-4357/abf3c2}, \href {https://ui.adsabs.harvard.edu/abs/2021ApJ...913...45O} {913, 45}

\bibitem[\protect\citeauthoryear{Paszke et~al.,}{Paszke et~al.}{2019}]{pytorch}
Paszke A.,  et~al., 2019, PyTorch: An Imperative Style, High-Performance Deep Learning Library (\mn@eprint {arXiv} {1912.01703})

\bibitem[\protect\citeauthoryear{Pedregosa et~al.,}{Pedregosa et~al.}{2011}]{sklearn}
Pedregosa F.,  et~al., 2011, Journal of Machine Learning Research, 12, 2825

\bibitem[\protect\citeauthoryear{{Peng} et~al.,}{{Peng} et~al.}{2010}]{Peng_2010}
{Peng} Y.-j.,  et~al., 2010, \mn@doi [\apj] {10.1088/0004-637X/721/1/193}, \href {https://ui.adsabs.harvard.edu/abs/2010ApJ...721..193P} {721, 193}

\bibitem[\protect\citeauthoryear{Peng, Maiolino  \& Cochrane}{Peng et~al.}{2015}]{pengStrangulationPrimaryMechanism2015}
Peng Y.,  Maiolino R.,   Cochrane R.,  2015, \mn@doi [Nature] {10.1038/nature14439}, 521, 192

\bibitem[\protect\citeauthoryear{{Pillepich} et~al.,}{{Pillepich} et~al.}{2018}]{TNG5}
{Pillepich} A.,  et~al., 2018, \mn@doi [\mnras] {10.1093/mnras/stx3112}, \href {https://ui.adsabs.harvard.edu/abs/2018MNRAS.475..648P} {475, 648}

\bibitem[\protect\citeauthoryear{{Planck Collaboration} et~al.,}{{Planck Collaboration} et~al.}{2016}]{Plank_2016}
{Planck Collaboration} et~al., 2016, \mn@doi [\aap] {10.1051/0004-6361/201525830}, \href {https://ui.adsabs.harvard.edu/abs/2016A&A...594A..13P} {594, A13}

\bibitem[\protect\citeauthoryear{{Rodighiero} et~al.,}{{Rodighiero} et~al.}{2011}]{rodighieroLESSERROLESTARBURSTS2011}
{Rodighiero} G.,  et~al., 2011, \mn@doi [\apjl] {10.1088/2041-8205/739/2/L40}, \href {https://ui.adsabs.harvard.edu/abs/2011ApJ...739L..40R} {739, L40}

\bibitem[\protect\citeauthoryear{{Rodr{\'\i}guez-Puebla}, {Primack}, {Avila-Reese}  \& {Faber}}{{Rodr{\'\i}guez-Puebla} et~al.}{2017}]{2017MNRAS.470..651R}
{Rodr{\'\i}guez-Puebla} A.,  {Primack} J.~R.,  {Avila-Reese} V.,   {Faber} S.~M.,  2017, \mn@doi [\mnras] {10.1093/mnras/stx1172}, \href {https://ui.adsabs.harvard.edu/abs/2017MNRAS.470..651R} {470, 651}

\bibitem[\protect\citeauthoryear{Schaefer et~al.,}{Schaefer et~al.}{2018}]{ram_2}
Schaefer A.~L.,  et~al., 2018, \mn@doi [\mnras] {10.1093/mnras/sty3258}, 483, 2851

\bibitem[\protect\citeauthoryear{{Schreiber, C.} et~al.,}{{Schreiber, C.} et~al.}{2015}]{Schreiber2015}
{Schreiber, C.} et~al., 2015, \mn@doi [A&A] {10.1051/0004-6361/201425017}, 575, A74

\bibitem[\protect\citeauthoryear{{Sheth} \& {Tormen}}{{Sheth} \& {Tormen}}{2004}]{Sheth_2004}
{Sheth} R.~K.,  {Tormen} G.,  2004, \mn@doi [\mnras] {10.1111/j.1365-2966.2004.07733.x}, \href {https://ui.adsabs.harvard.edu/abs/2004MNRAS.350.1385S} {350, 1385}

\bibitem[\protect\citeauthoryear{{Sousbie}}{{Sousbie}}{2011}]{Sousbie_2011}
{Sousbie} T.,  2011, \mn@doi [\mnras] {10.1111/j.1365-2966.2011.18394.x}, \href {https://ui.adsabs.harvard.edu/abs/2011MNRAS.414..350S} {414, 350}

\bibitem[\protect\citeauthoryear{{Springel} et~al.,}{{Springel} et~al.}{2018}]{TNG1}
{Springel} V.,  et~al., 2018, \mn@doi [\mnras] {10.1093/mnras/stx3304}, \href {https://ui.adsabs.harvard.edu/abs/2018MNRAS.475..676S} {475, 676}

\bibitem[\protect\citeauthoryear{{The pandas development team}}{{The pandas development team}}{2020}]{reback2020pandas}
{The pandas development team} 2020, \mn@doi{10.5281/zenodo.3509134}

\bibitem[\protect\citeauthoryear{{Tinker}, {Conroy}, {Norberg}, {Patiri}, {Weinberg}  \& {Warren}}{{Tinker} et~al.}{2008}]{Tinker_2008}
{Tinker} J.~L.,  {Conroy} C.,  {Norberg} P.,  {Patiri} S.~G.,  {Weinberg} D.~H.,   {Warren} M.~S.,  2008, \mn@doi [\apj] {10.1086/589983}, \href {https://ui.adsabs.harvard.edu/abs/2008ApJ...686...53T} {686, 53}

\bibitem[\protect\citeauthoryear{{Villaescusa-Navarro} et~al.,}{{Villaescusa-Navarro} et~al.}{2021}]{Villaescusa-Navarro2021}
{Villaescusa-Navarro} F.,  et~al., 2021, \mn@doi [\apj] {10.3847/1538-4357/abf7ba}, \href {https://ui.adsabs.harvard.edu/abs/2021ApJ...915...71V} {915, 71}

\bibitem[\protect\citeauthoryear{Virtanen et~al.,}{Virtanen et~al.}{2020}]{SciPy}
Virtanen P.,  et~al., 2020, \mn@doi [Nature Methods] {10.1038/s41592-019-0686-2}, \href {https://rdcu.be/b08Wh} {17, 261}

\bibitem[\protect\citeauthoryear{Wang \& White}{Wang \& White}{2012}]{1halo_theori}
Wang W.,  White S. D.~M.,  2012, \mn@doi [\mnras] {10.1111/j.1365-2966.2012.21256.x}, 424, 2574

\bibitem[\protect\citeauthoryear{{Wang}, {Mo}  \& {Jing}}{{Wang} et~al.}{2007}]{Wang_2007}
{Wang} H.~Y.,  {Mo} H.~J.,   {Jing} Y.~P.,  2007, \mn@doi [\mnras] {10.1111/j.1365-2966.2006.11316.x}, \href {https://ui.adsabs.harvard.edu/abs/2007MNRAS.375..633W} {375, 633}

\bibitem[\protect\citeauthoryear{{Wang}, {Avestruz}, {Guo}, {Wang}  \& {Wang}}{{Wang} et~al.}{2024a}]{Wang_2024}
{Wang} K.,  {Avestruz} C.,  {Guo} H.,  {Wang} W.,   {Wang} P.,  2024a, \mn@doi [\mnras] {10.1093/mnras/stae1805}, \href {https://ui.adsabs.harvard.edu/abs/2024MNRAS.532.4616W} {532, 4616}

\bibitem[\protect\citeauthoryear{{Wang} et~al.,}{{Wang} et~al.}{2024b}]{WangY_2024}
{Wang} Y.,  et~al., 2024b, \mn@doi [\apj] {10.3847/1538-4357/ad7f4c}, \href {https://ui.adsabs.harvard.edu/abs/2024ApJ...976..119W} {976, 119}

\bibitem[\protect\citeauthoryear{Wechsler \& Tinker}{Wechsler \& Tinker}{2018}]{annurev:galaxy-halo_connction}
Wechsler R.~H.,  Tinker J.~L.,  2018, \mn@doi [\araa] {https://doi.org/10.1146/annurev-astro-081817-051756}, 56, 435

\bibitem[\protect\citeauthoryear{{Wechsler}, {Bullock}, {Primack}, {Kravtsov}  \& {Dekel}}{{Wechsler} et~al.}{2002}]{Wechsler_2002}
{Wechsler} R.~H.,  {Bullock} J.~S.,  {Primack} J.~R.,  {Kravtsov} A.~V.,   {Dekel} A.,  2002, \mn@doi [\apj] {10.1086/338765}, \href {https://ui.adsabs.harvard.edu/abs/2002ApJ...568...52W} {568, 52}

\bibitem[\protect\citeauthoryear{Weinmann, {van den Bosch}, Yang  \& Mo}{Weinmann et~al.}{2006}]{weinmannPropertiesGalaxyGroups2006}
Weinmann S.~M.,  {van den Bosch} F.~C.,  Yang X.,   Mo H.~J.,  2006, \mn@doi [\mnras] {10.1111/j.1365-2966.2005.09865.x}, 366, 2

\bibitem[\protect\citeauthoryear{{White} \& {Rees}}{{White} \& {Rees}}{1978}]{1978MNRAS.183..341W}
{White} S.~D.~M.,  {Rees} M.~J.,  1978, \mn@doi [\mnras] {10.1093/mnras/183.3.341}, \href {https://ui.adsabs.harvard.edu/abs/1978MNRAS.183..341W} {183, 341}

\bibitem[\protect\citeauthoryear{{Woo} et~al.,}{{Woo} et~al.}{2013}]{Woo_2013}
{Woo} J.,  et~al., 2013, \mn@doi [\mnras] {10.1093/mnras/sts274}, \href {https://ui.adsabs.harvard.edu/abs/2013MNRAS.428.3306W} {428, 3306}

\bibitem[\protect\citeauthoryear{Wu, Jespersen  \& Wechsler}{Wu et~al.}{2024}]{Wu_2024}
Wu J.~F.,  Jespersen C.~K.,   Wechsler R.~H.,  2024, \mn@doi [\apj] {10.3847/1538-4357/ad7bb3}, 976, 37

\bibitem[\protect\citeauthoryear{Yuan et~al.,}{Yuan et~al.}{2025}]{yuanUnravelingEmissionLine2025}
Yuan S.,  et~al., 2025, \mn@doi [\mnras] {10.1093/mnras/staf368}, 538, 1216

\bibitem[\protect\citeauthoryear{{Zehavi}, {Contreras}, {Padilla}, {Smith}, {Baugh}  \& {Norberg}}{{Zehavi} et~al.}{2018}]{Zehavi_2018}
{Zehavi} I.,  {Contreras} S.,  {Padilla} N.,  {Smith} N.~J.,  {Baugh} C.~M.,   {Norberg} P.,  2018, \mn@doi [\apj] {10.3847/1538-4357/aaa54a}, \href {https://ui.adsabs.harvard.edu/abs/2018ApJ...853...84Z} {853, 84}

\bibitem[\protect\citeauthoryear{{Zhang}, {Wang}, {Zhang}, {Sun}, {He}, {Contardo}, {Villaescusa-Navarro}  \& {Ho}}{{Zhang} et~al.}{2019}]{Zhang_2019}
{Zhang} X.,  {Wang} Y.,  {Zhang} W.,  {Sun} Y.,  {He} S.,  {Contardo} G.,  {Villaescusa-Navarro} F.,   {Ho} S.,  2019, \mn@doi [arXiv e-prints] {10.48550/arXiv.1902.05965}, \href {https://ui.adsabs.harvard.edu/abs/2019arXiv190205965Z} {}

\makeatother
\end{thebibliography}




\appendix
\section*{Appendices}
\renewcommand{\thesection}{Appendix \Alph{section}}
\renewcommand{\thefigure}{A\arabic{figure}}
\renewcommand{\thetable}{A\arabic{table}}

\subsection{Dataset used} \label{sec:data_app}
\begin{figure}
    \centering
    \begin{minipage}{0.24\textwidth}
        \centering
        \includegraphics[width=\linewidth]{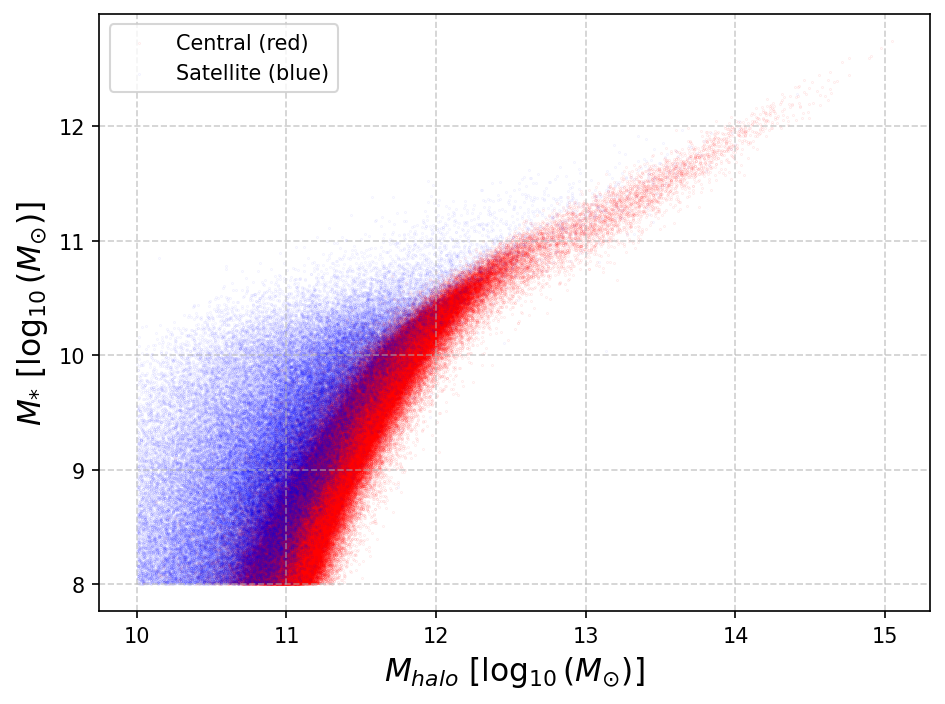}
    \end{minipage}
    \hfill
    \begin{minipage}{0.23\textwidth}
        \centering
        \includegraphics[width=\linewidth]{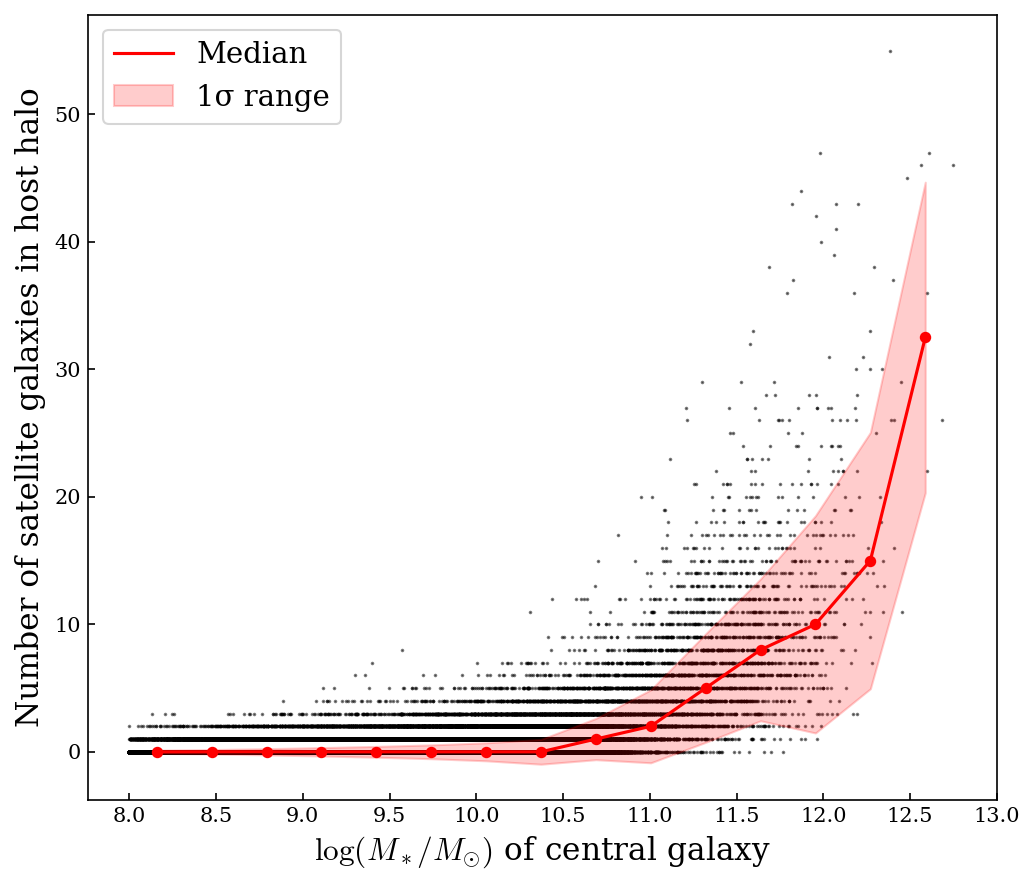}
    \end{minipage}
    \caption{
    Left panel: The distribution of central (291,885 objects) and satellite (101,846 objects) galaxies in TNG300-1 ($z=0$) used in this study.
    Right panel: The number of satellite galaxies hosted by central galaxies in the dataset. The median and $1\sigma$ regions in each bin are shown.
    }
    \label{Fig.halo-stellarmass_relation}
\end{figure}

\begin{figure}
    \centering
    \includegraphics[width=\linewidth]{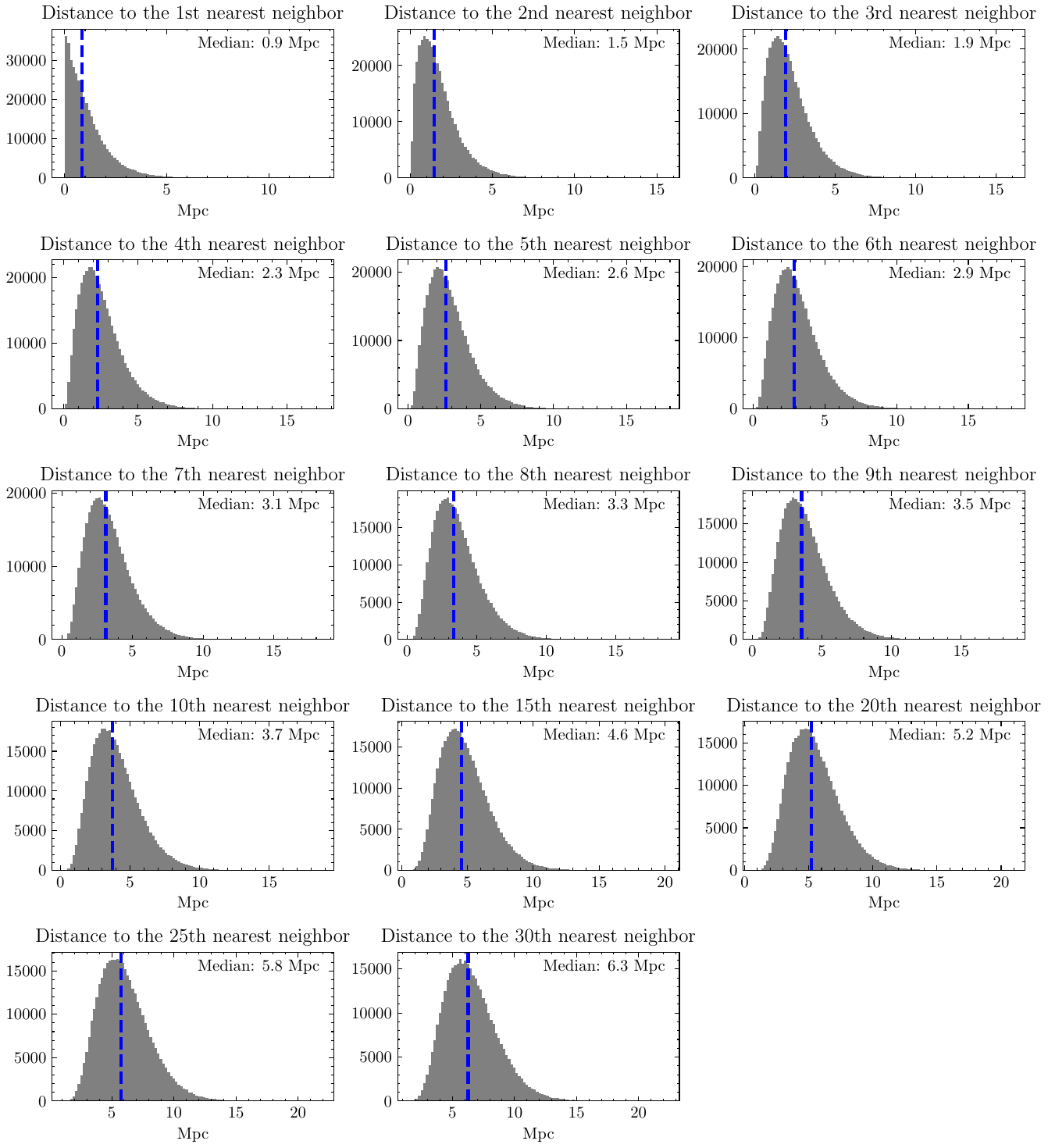}
    \caption{Distribution of distances to the $n$th nearest neighbor galaxies. Each panel presents a histogram of distances to the $n$th nearest neighbor, with $n$ varying across panels. The blue dashed lines represent the median distances, whose values are displayed in the upper right corner of each panel.}
    \label{Fig.Distri_nth}
\end{figure}

The distribution of all datasets used in this study is shown in Fig.~\ref{Fig.halo-stellarmass_relation}, and ~\ref{Fig.Distri_nth}.

\subsection{Difference between with and without environmental factors, keeping $V_{\mathrm{max}}$ and $r_{50}$ as input features} \label{sec:appe_diffe_with_without}
\begin{figure}
    \centering
    \begin{subfigure}{0.24\textwidth}
        \centering
        \includegraphics[width=\linewidth]{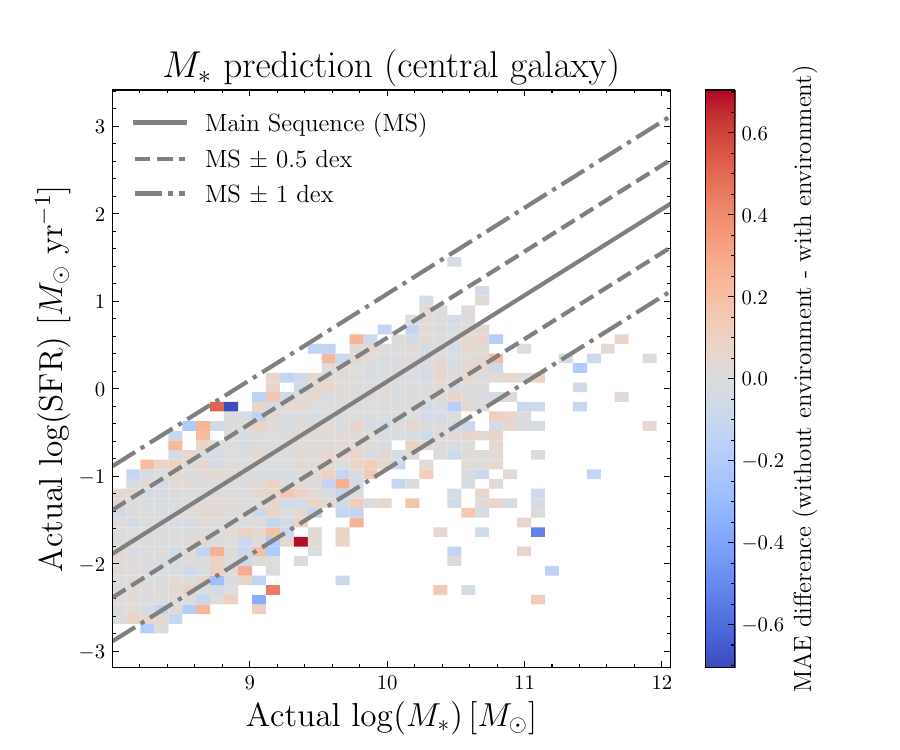}
    \end{subfigure}%
    \hfill
    \begin{subfigure}{0.24\textwidth}
        \centering
        \includegraphics[width=\linewidth]{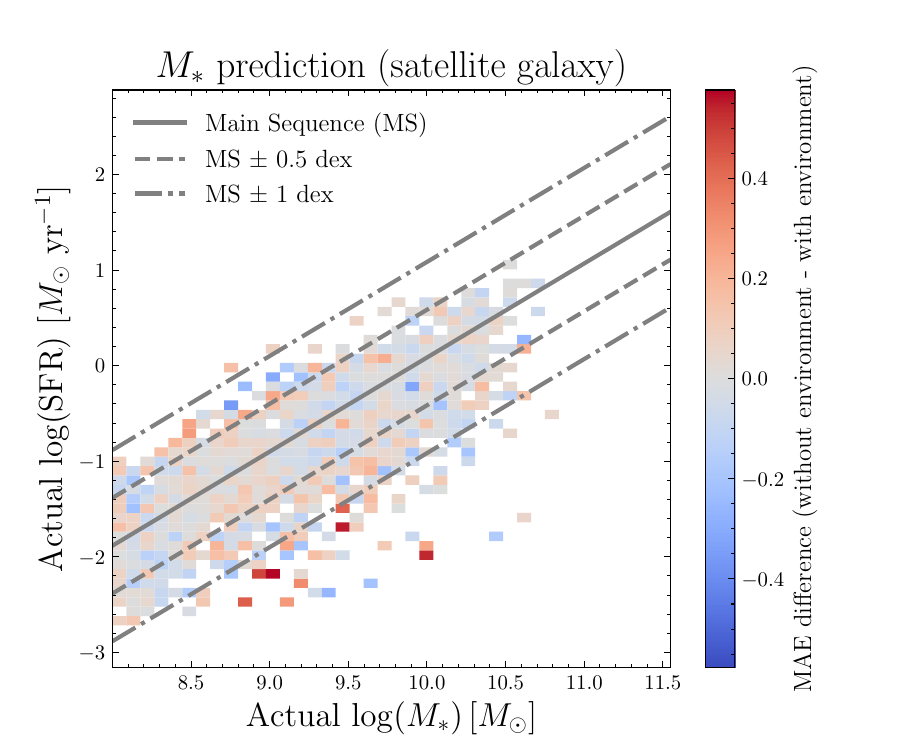}
    \end{subfigure}
    \hfill
    \begin{subfigure}{0.24\textwidth}
        \centering
        \includegraphics[width=\linewidth]{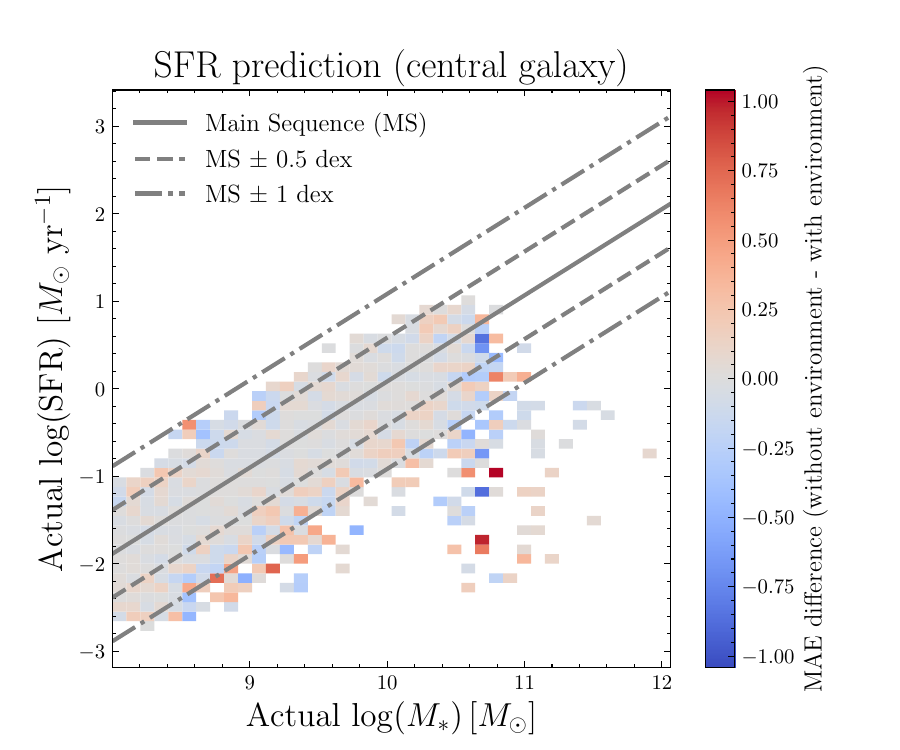}
    \end{subfigure}%
    \hfill
    \begin{subfigure}{0.24\textwidth}
        \centering
        \includegraphics[width=\linewidth]{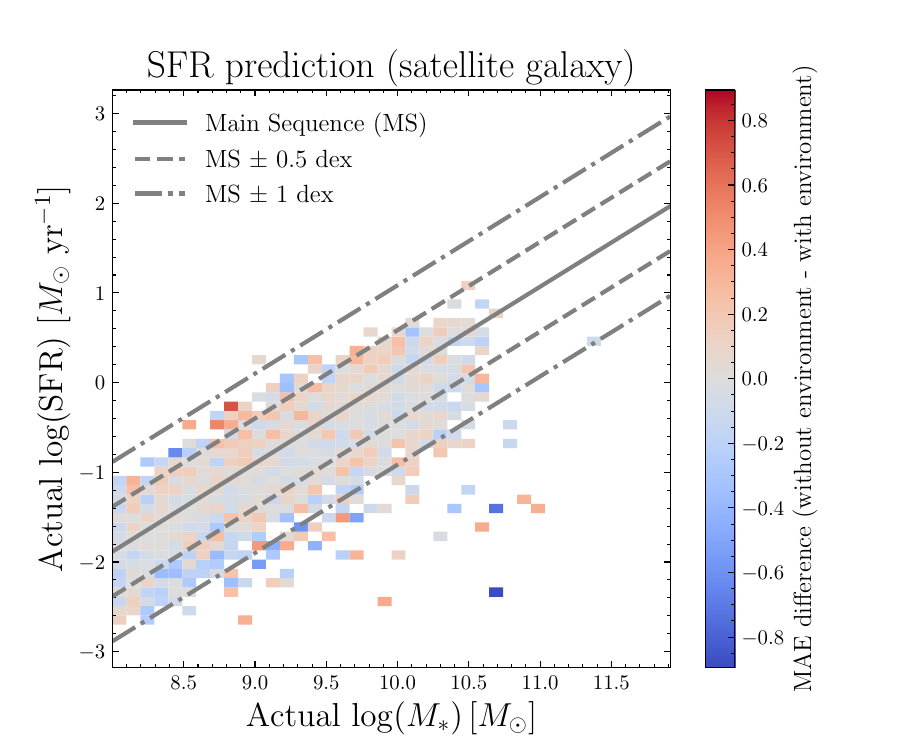}
    \end{subfigure}
    \caption{Difference in MAE with and without environment. Red in the color bar means that using the environment improves prediction accuracy. Blue is the opposite, meaning worse accuracy.}
    \label{Fig.MAE_diff}
\end{figure}

Fig.~\ref{Fig.MAE_diff} show how much the prediction accuracy improves for the one-neighbor model (for $M_*$ prediction) and the three-neighbors model (for SFR prediction) compared to the zero-neighbor model (without environmental information), as presented in Section.~\ref{sec:result_model}.

An analysis of the changes in MAE due to the inclusion of environmental information reveal that, while the overall prediction accuracy tended to improve, there are specific regions where MAE increases instead.  
In particular, when the data are divided based on $M_*$ and SFR, the inclusion of environmental information led to greater variability in predictions, resulting in a tendency for MAE to increase.  
One possible explanation for this phenomenon is that the increased number of input parameters makes optimization more challenging. Adding environmental information increases the dimensionality of the features the model must consider, potentially making it harder to find an optimal solution.  
Additionally, statistical fluctuations due to an insufficient number of samples may also play a role. 
In particular, galaxies that deviate from the Main Sequence are relatively rare, and the limited data available for such galaxies can lead to increased statistical fluctuations. As a result, in regions where these samples are present, the inclusion of environmental information may lead to a higher MAE.  
Furthermore, the relationship between environmental factors and galaxy properties is not necessarily straightforward, which may also contribute to the increase in MAE. While adding environmental information can enhance the model's expressiveness, in some cases, it may introduce unnecessary noise, ultimately reducing prediction accuracy.  
For example, if a massive halo is present near a given galaxy, the model may be more likely to predict that the galaxy is quiescent. However, some galaxies continue to form stars despite their environment, and incorporating environmental information could, in such cases, lead to a decrease in prediction accuracy.  

Thus, while environmental information is useful for capturing the average properties of galaxies, it does not necessarily improve the prediction accuracy for individual galaxies.

\subsection{Prediction of $M_*$ and SFR excluding $V_{\mathrm{max}}$ and $r_{50}$ from input features} \label{sec:appe_predic_no-assembly}
\begin{figure}
    \centering
    \begin{minipage}{0.21\textwidth}
        \centering
        \includegraphics[width=\linewidth]{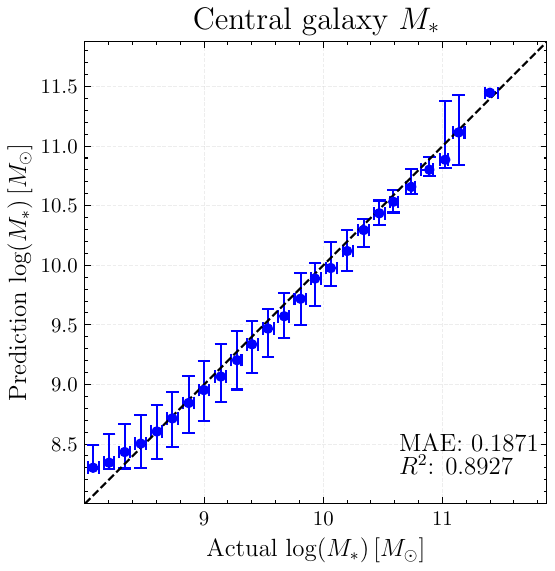}
    \end{minipage}
    \begin{minipage}{0.241\textwidth}
        \centering
        \includegraphics[width=\linewidth]{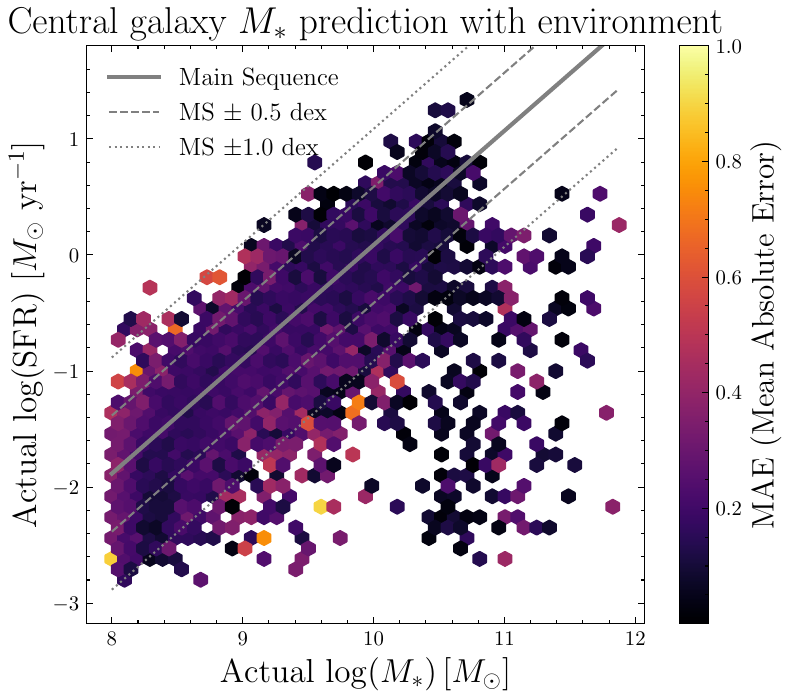}
    \end{minipage}
    \vspace{5mm}
    \begin{minipage}{0.21\textwidth}
        \centering
        \includegraphics[width=\linewidth]{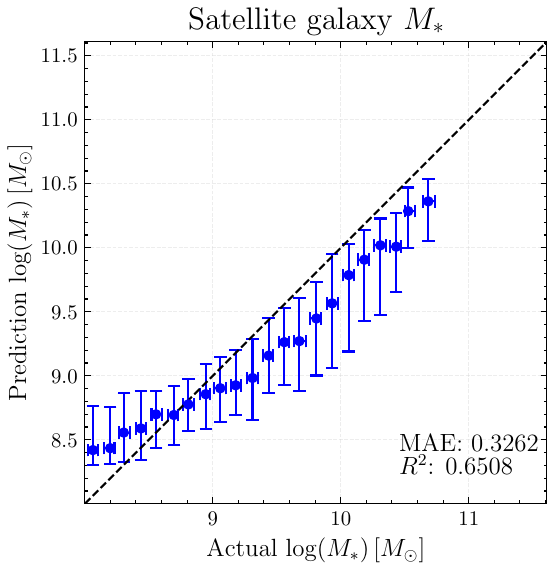}
    \end{minipage}
    \begin{minipage}{0.241\textwidth}
        \centering
        \includegraphics[width=\linewidth]{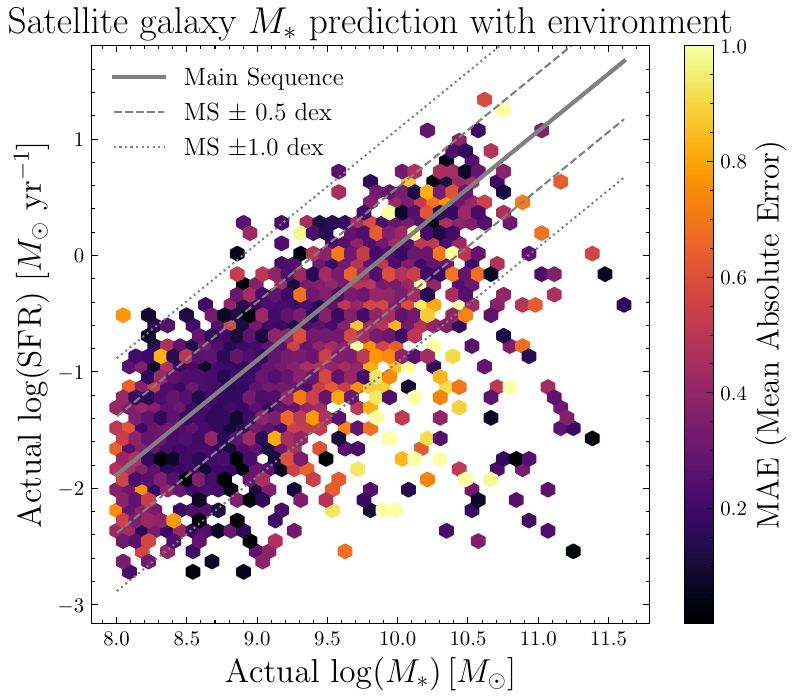}
    \end{minipage}
    \caption{$M_*$ prediction result of one-neighbor model. $V_{\mathrm{max}}$ and $r_{50}$ are excluded from the input features.}
    \label{Fig.pre_sm_no_1}
\end{figure}

\begin{figure}
    \centering
    \begin{minipage}{0.21\textwidth}
        \centering
        \includegraphics[width=\linewidth]{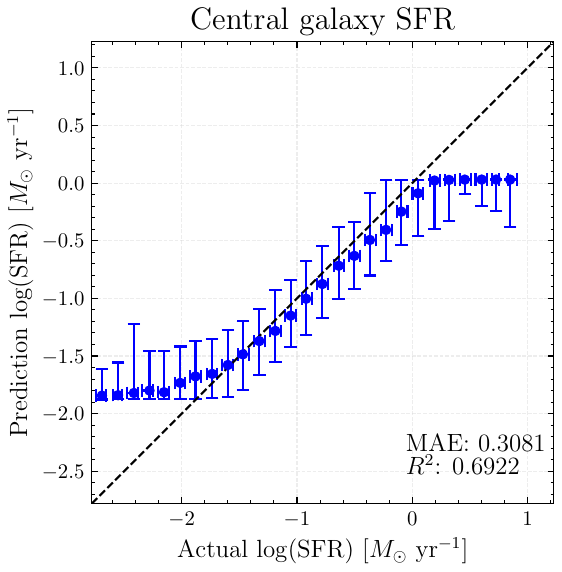}
    \end{minipage}
    \begin{minipage}{0.241\textwidth}
        \centering
        \includegraphics[width=\linewidth]{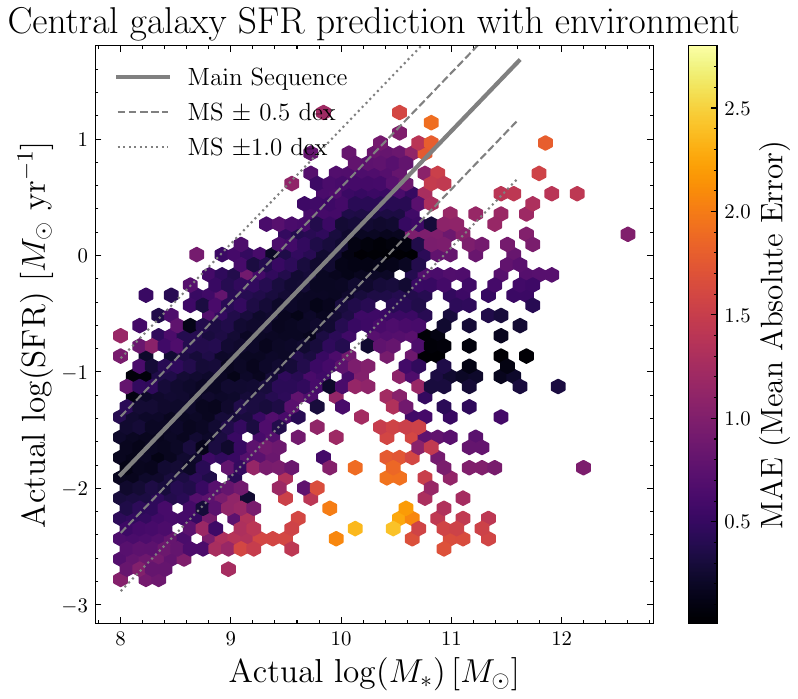}
    \end{minipage}
    \vspace{5mm}
    \begin{minipage}{0.21\textwidth}
        \centering
        \includegraphics[width=\linewidth]{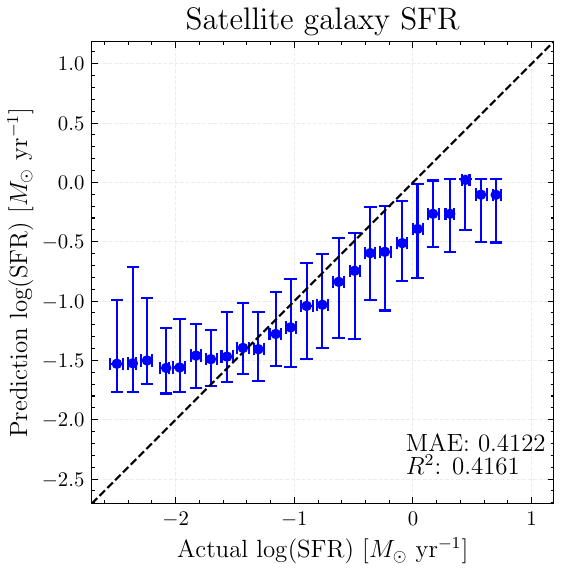}
    \end{minipage}
    \begin{minipage}{0.241\textwidth}
        \centering
        \includegraphics[width=\linewidth]{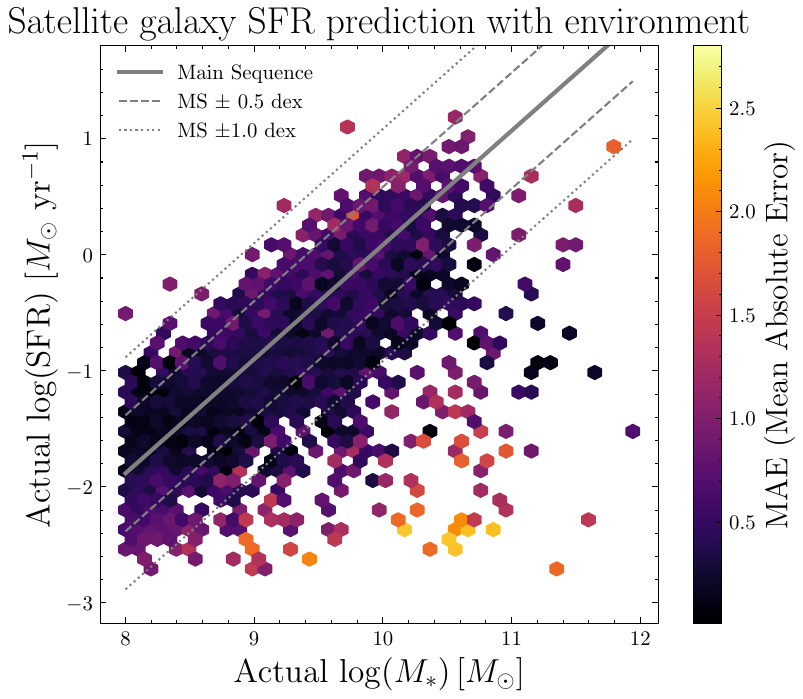}
    \end{minipage}
    \caption{SFR prediction result of three-neighbors model. $V_{\mathrm{max}}$ and $r_{50}$ are excluded from the input features.}
    \label{Fig.pre_sfr_no_3}
\end{figure}

\begin{figure*}
    \centering
    \begin{subfigure}{0.5\textwidth}
        \centering
        \includegraphics[width=0.87 \textwidth]{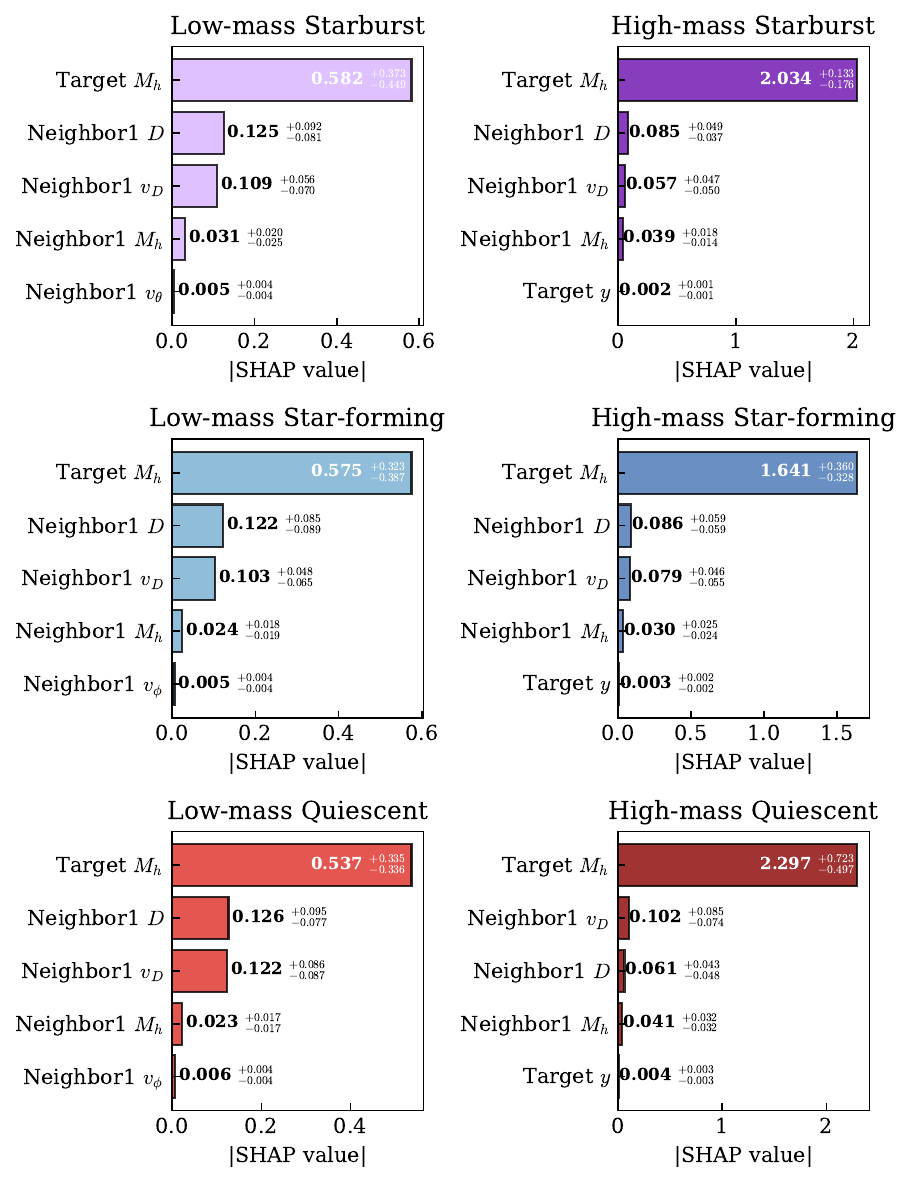}
        \caption{Predictions for central galaxies}
    \end{subfigure}%
    \hfill
    \begin{subfigure}{0.5\textwidth}
        \centering
        \includegraphics[width=0.87\textwidth]{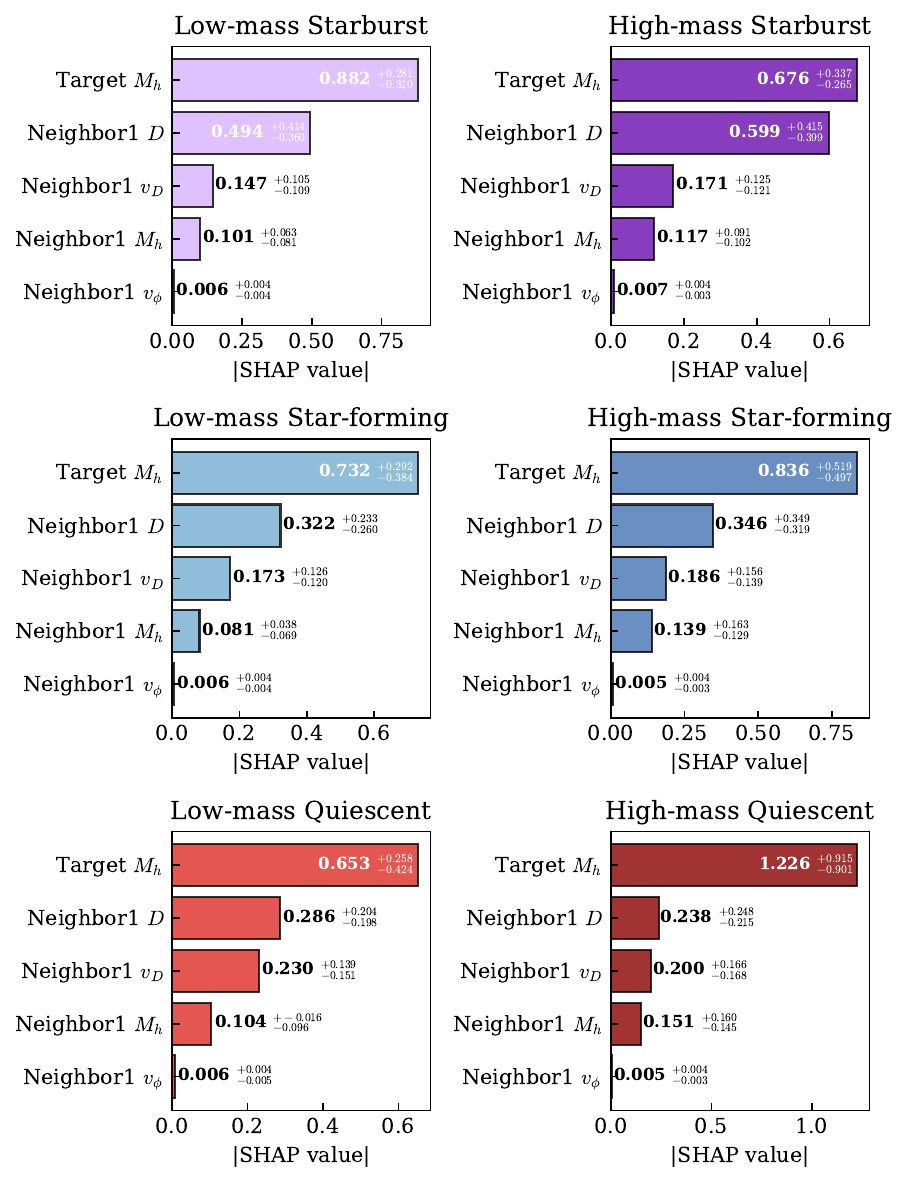}
        \caption{Predictions for satellite galaxies}
    \end{subfigure}
    \caption{Top five features contributing to $M_*$ predictions for pure environmental effects of the one-neighbor model. The values in the bars represent the mean absolute SHAP value and the 16th and 84th percentiles.}
    \label{Fig.SHAP_summary_SM_no}
\end{figure*}

\begin{figure*}
    \centering
    \begin{subfigure}{0.5\textwidth}
        \centering
        \includegraphics[width=0.87\textwidth]{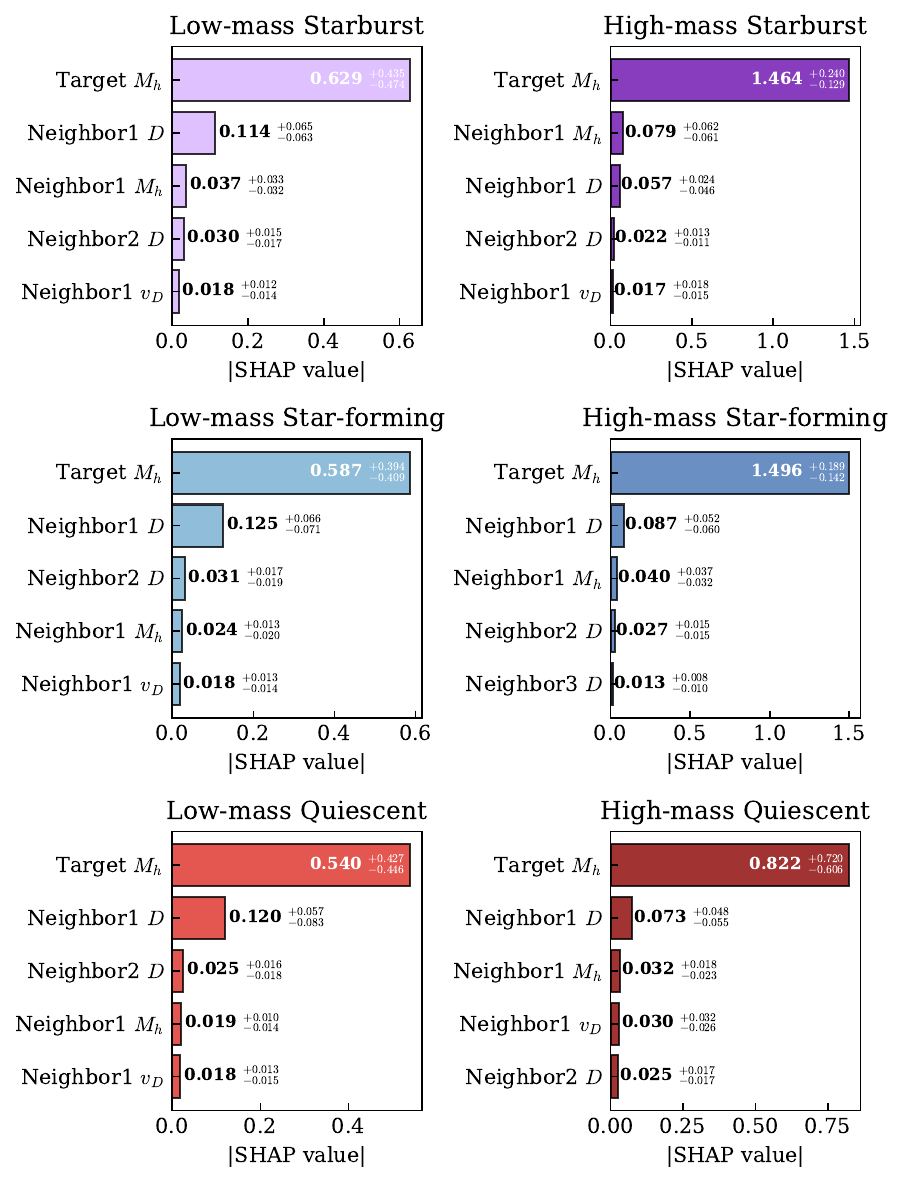}
        \caption{Predictions for central galaxies}
    \end{subfigure}%
    \hfill
    \begin{subfigure}{0.5\textwidth}
        \centering
        \includegraphics[width=0.87\textwidth]{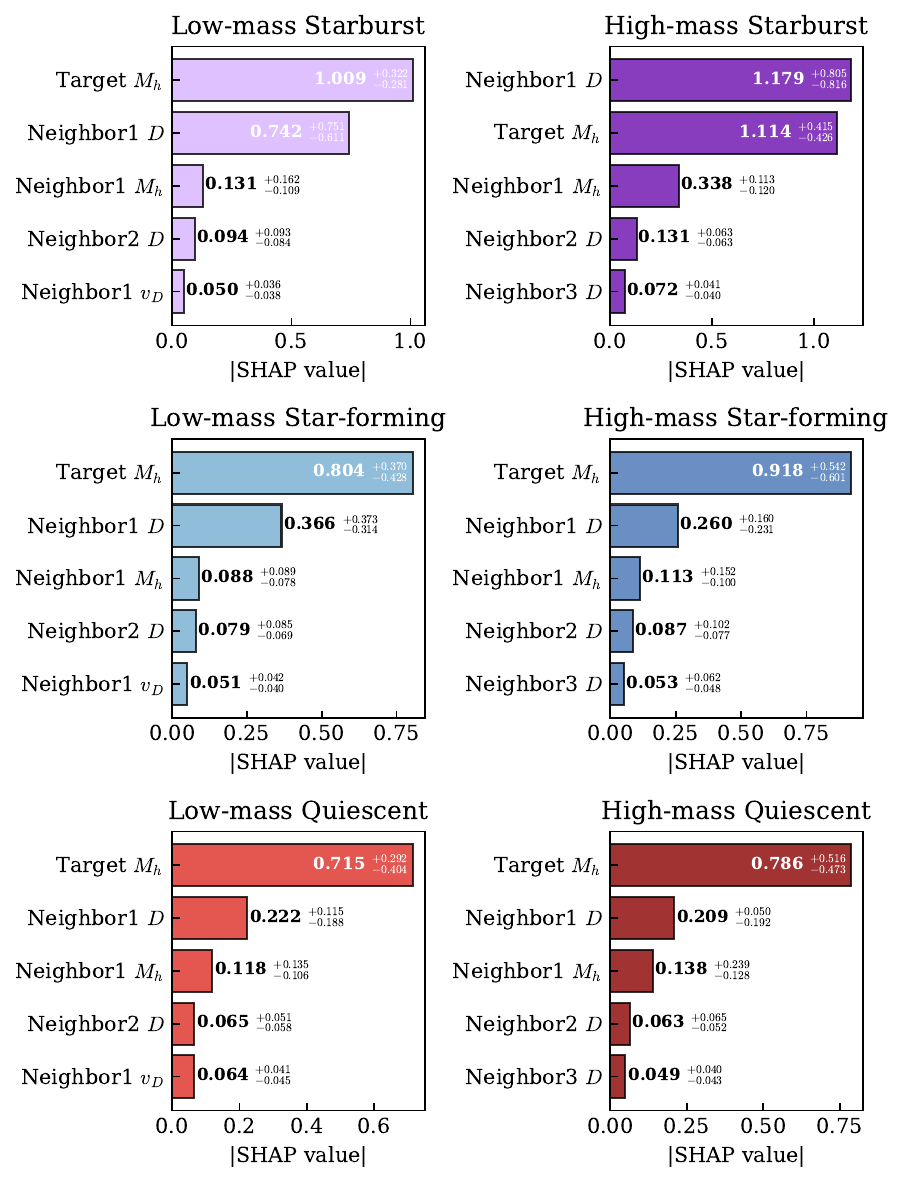}
        \caption{Predictions for satellite galaxies}
    \end{subfigure}
    \caption{Top five features contributing to SFR predictions for pure environmental effects of the three-neighbors model.}
    \label{Fig.SHAP_summary_SFR_no}
\end{figure*}

The prediction accuracy of the model used in Section~\ref{sec:discuss_no-assembly} is shown in Fig.\ref{Fig.pre_sm_no_1}, and \ref{Fig.pre_sfr_no_3}.
Although the results of the SHAP analysis for this model are presented in Fig.~\ref{tab:shap_frac_no} and Fig.~\ref{Fig.SHAP_feature_no}, more detailed results are provided in Fig.\ref{Fig.SHAP_summary_SM_no}, and \ref{Fig.SHAP_summary_SFR_no}. 

\begin{figure*}
    \centering
    \begin{minipage}{0.69\textwidth}
        \centering
        \includegraphics[width=\linewidth]{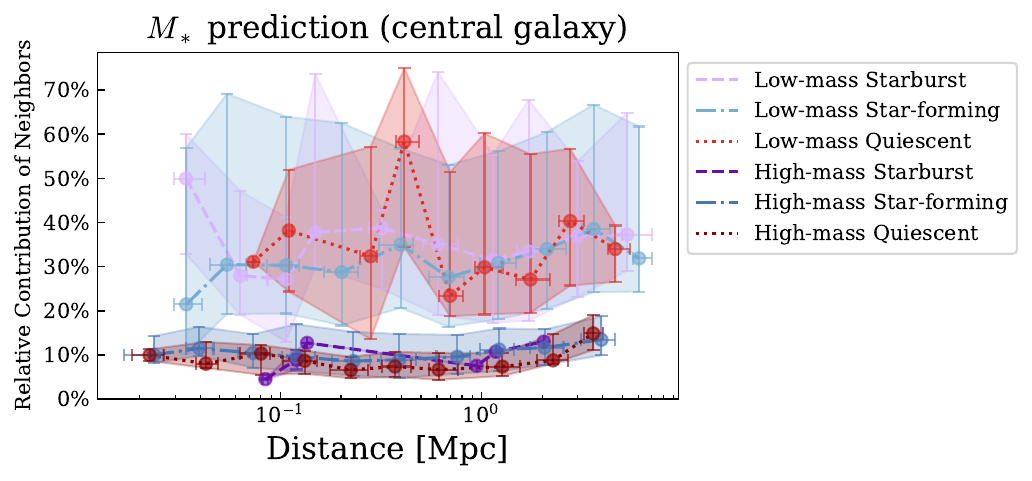}
    \end{minipage}
    \begin{minipage}{0.69\textwidth}
        \centering
        \includegraphics[width=\linewidth]{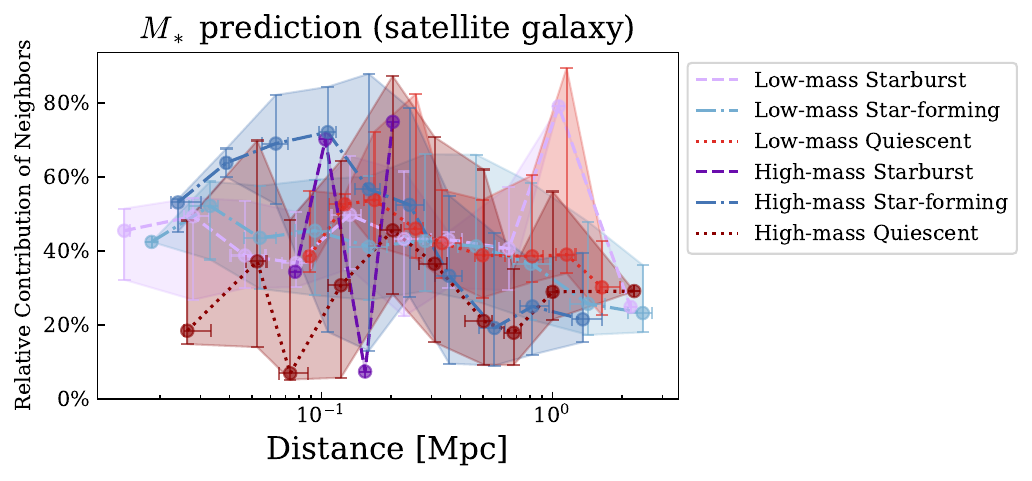}
    \end{minipage}
    \begin{minipage}{0.69\textwidth}
        \centering
        \includegraphics[width=\linewidth]{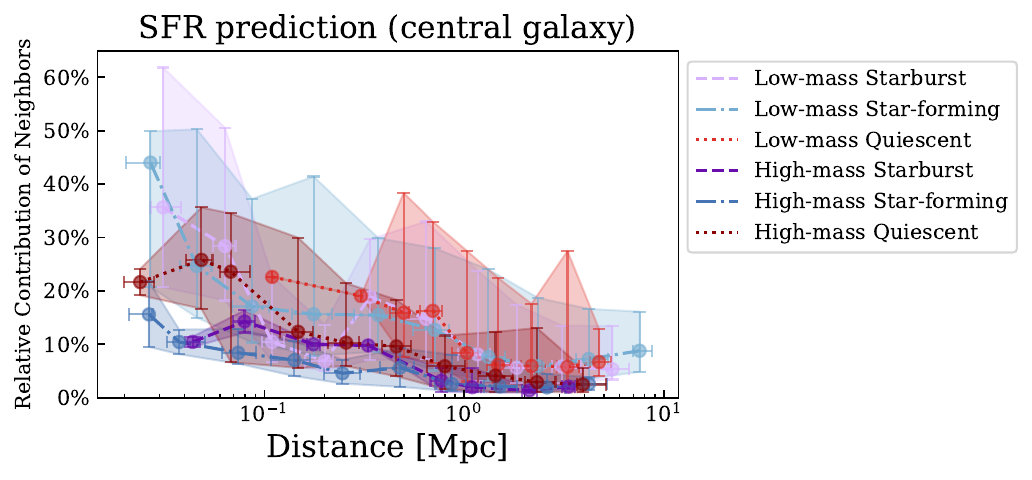}
    \end{minipage}
    \begin{minipage}{0.69\textwidth}
        \centering
        \includegraphics[width=\linewidth]{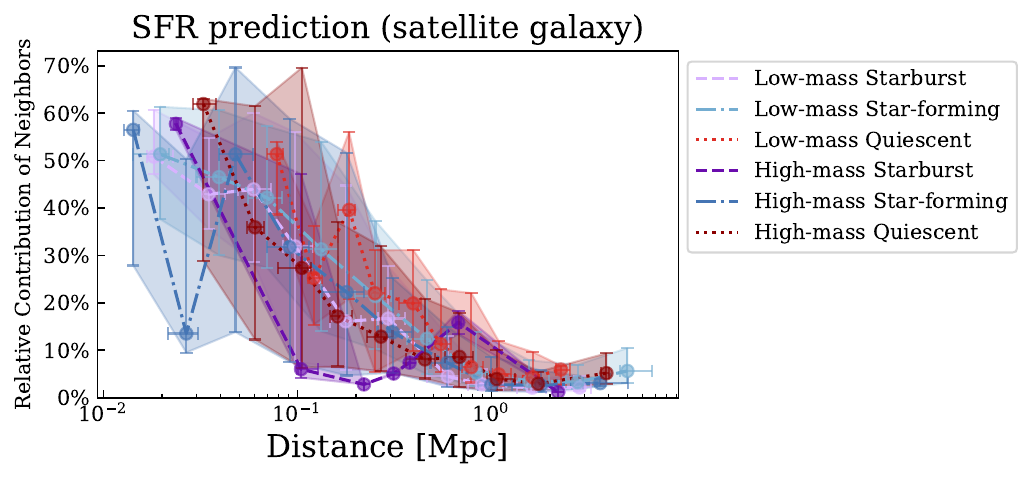}
    \end{minipage}
    \caption{
       The dependence of the SHAP values of environmental features on distance. 
       $M_*$ prediction uses the one-neighbor model, and SFR prediction uses the three-neighbors model.
       The x-axis is divided into ten logarithmic bins.
       Symbols indicate median values in each bin; error bars represent the 16th and 84th percentiles. Bins without data are not shown.
    }
    \label{Fig.D_bin_no}
\end{figure*}

Furthermore, in Fig.~\ref{Fig.D_bin_no} we divide the nearby galaxies into uniform bins on a logarithmic scale and examine their respective contributions.
The variations in contribution are largely within the range of the error bars, with no clear distance dependence observed. One potential explanation for this is that the multi-layer perceptron (MLP) model requires equal input dimensions, leading to the definition of neighboring galaxies based on the $N$-nearest approach, which may introduce limitations. This remains an open issue for future investigation. Furthermore, our results suggest that environmental influences are not determined solely by simple distance scales; they may also be influenced by tidal fields and the filamentary structure of DM. 



\subsection{Distribution of neighbor distances grouped by first nearest neighbor distance} \label{sec:D_Q_app}
We analyze the distribution of distances to the 30 nearest neighbors for each galaxy, categorized into quartile groups based on the distance to the first nearest neighbor ($D_1$): Q1 ($D_1 \leq Q_{25}$), Q2 ($Q_{25} < D_1 \leq Q_{50}$), Q3 ($Q_{50} < D_1 \leq Q_{75}$), and Q4 ($D_1 > Q_{75}$). This classification allows us to examine how proximity to the first nearest neighbor affects the distribution of subsequent neighbor distances.

\begin{figure*}
    \centering
    \includegraphics[width=0.68\textwidth]{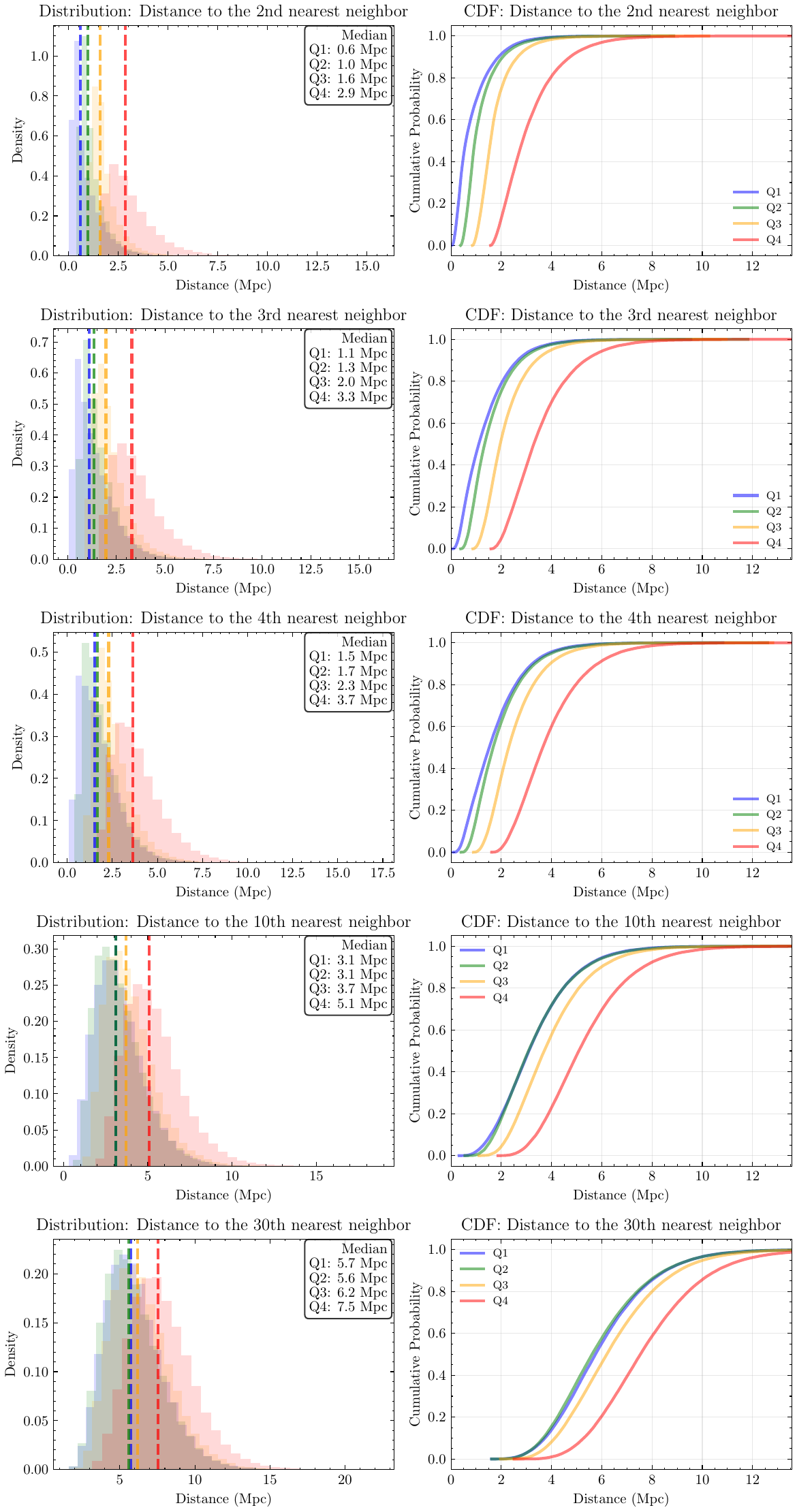}
    \caption{
    Distributions (left panel) and cumulative distribution functions (CDFs, right panel) of the distances to the 2nd to 30th nearest neighbors ($D_2, D_3, D_4, D_{10}, D_{30}$), grouped by the quartile of the first nearest neighbor distance ($D_1$). Each line represents a quartile group—Q1 (closest $D_1$) to Q4 (farthest $D_1$). Histograms show normalized densities, and dashed lines indicate the median values.}
    \label{Fig.Distri_Q}
\end{figure*}

Fig.~\ref{Fig.Distri_Q} presents the distributions and cumulative distribution functions (CDFs) for $D_2, D_3, D_4, D_{10}$ and $D_{30}$.
The histograms reveal significant differences in median values across quartile groups for each neighbor distance (e.g., for $ D_2 $: Q1: 0.6 Mpc, Q4: 2.9 Mpc).
From $D_2$ to $D_4$, the Q1 distributions are left-shifted and more concentrated, whereas Q4 distributions are broader with higher medians.
This trend persists for $ D_{10} $ and $ D_{30} $, where the Q4 distribution exhibits a longer tail (e.g., for $ D_{30} $: Q1: 5.7 Mpc, Q4: 7.5 Mpc).

The CDFs clearly visualize the cumulative probability differences among groups. 
For $D_2$, the Q1 CDF rises sharply around 0.5 Mpc, whereas the Q4 CDF increases gradually up to 1.5 Mpc, indicating a substantial distribution difference. 
The distribution difference narrows toward $ D_{30} $, though the Q4 CDF remains right-shifted. 
Kolmogorov-Smirnov (KS) tests confirmed statistically significant differences for all quartile pairs with $p < 0.01 $.

These findings suggest that the $D_1$ is correlated with the distribution of subsequent nearest neighbor distances, with closer objects (Q1) consistently exhibiting compact distributions and farther objects (Q4) showing more dispersed patterns.
Based on these results, it can be inferred that when the first nearest neighbor is close, the subsequent neighbors also tend to be nearby rather than widely dispersed, suggesting a clustering tendency.

\bsp	
\label{lastpage}
\end{document}